\documentclass[11pt,twoside,a4paper,cmspaper,final,collab]{cms-tdr}
\def\svnVersion{8361506-D}\def\svnDate{2026/09/23}\def\cmsCernNoTag{CERN-EP-2026-252}\def\cmsCernDate{\today}\def\cmsMessage{Submitted to the Journal of High Energy Physics}
\begin{document}\cmsNoteHeader{SMP-24-010}

\newcommand{\TUnfold}{\textsc{TUnfold}\xspace}
\newcommand{\NNLOJet}{\textsc{nnlojet}\xspace}
\newcommand{\FastNLO}{\textsc{fastNLO}\xspace}
\newcommand{\APPLfast}{\textsc{APPLfast}\xspace}
\newcommand{\HERWIGvii}{{\HERWIG}\,7\xspace}
\newcommand{\minnlops}{\textsc{MiNNLOPS}\xspace}
\newcommand{\mz}{\ensuremath{m_{\PZ}}\xspace}
\newcommand{\ptz}{\ensuremath{\pt^{\PZ}}\xspace}
\newcommand{\yjet}{\ensuremath{y_{\text{jet}}}\xspace}
\newcommand{\yz}{\ensuremath{y_{\PZ}}\xspace}
\newcommand{\yboost}{\ensuremath{y_{\text{b}}}\xspace}
\newcommand{\ystar}{\ensuremath{y^\ast}\xspace}
\newcommand{\ybys}{\ensuremath{(\yboost,\ystar)}\xspace}
\newcommand{\as}{\alpS}
\newcommand{\asmz}{\ensuremath{\as(\mz)}\xspace}
\newcommand{\mur}{\ensuremath{\mu_{\mathrm{R}}}\xspace}
\newcommand{\muf}{\ensuremath{\mu_{\mathrm{F}}}\xspace}
\newcommand{\mun}{\ensuremath{\mu_0}\xspace}
\newcommand{\ord}{\ensuremath{\mathcal{O}}\xspace}
\newcommand{\zj}{\ensuremath{\PZ/\PGgst{+}\text{jet}}\xspace}
\newcommand{\zmmj}{\ensuremath{\PZ/\PGgst(\to\PGm\PGm){+}\text{jet}}\xspace}
\newcommand{\zmmjx}{\ensuremath{\zmmj+\PX}\xspace}
\newcommand{\pp}{\ensuremath{\Pp\Pp}\xspace}
\newcommand{\mZ}{\ensuremath{m_{\PZ}}\xspace}
\newcommand{\sigLO}{\ensuremath{\sigma^{\text{LO}}}\xspace}
\newcommand{\sigQCDNLO}{\ensuremath{\sigma^{\text{NLO}}_{\text{QCD}}}\xspace}
\newcommand{\sigEWNLO}{\ensuremath{\sigma^{\text{NLO}}_{\text{EW}}}\xspace}
\newcommand{\sigQCDNNLO}{\ensuremath{\sigma^{\text{NNLO}}_{\text{QCD}}}\xspace}

\cmsNoteHeader{SMP-24-010}
\title{Measurement of the triple-differential cross section of \texorpdfstring{{\PZ}+jet}{Z+jet} production in proton-proton collisions at \texorpdfstring{$\sqrt{s}=13\TeV$}{sqrt(s)=13 TeV}}

\author[cern]{The CMS Collaboration}
\date{\today}

\abstract{A measurement is presented of the triple-differential production cross section of \zmmj events using proton-proton collision data recorded at a center-of-mass energy of 13\TeV by the CMS experiment at the CERN LHC. The data, collected in the years 2016--2018, correspond to an integrated luminosity of 138\fbinv. The cross section is measured as a function of the transverse momentum of the muon pair, the boost in rapidity of the center-of-mass system of the muon pair and the leading jet with respect to the laboratory frame, and half of their absolute rapidity separation. This choice of observables ensures sensitivity to the scattering angle in the center-of-mass frame and the fractional momenta of the interacting partons.  After unfolding for detector effects in all three dimensions simultaneously, the results are compared with predictions at next-to-next-to-leading order in perturbative quantum chromodynamics corrected for electroweak and nonperturbative effects. This measurement provides valuable information towards characterizing the partonic structure of the proton and determining the strong coupling.}

\hypersetup{%
pdfauthor={CMS Collaboration},%
pdftitle={Measurement of the triple-differential cross section of Z+jet production in proton-proton collisions at sqrt(s)=13 TeV},%
pdfsubject={CMS},%
pdfkeywords={CMS, QCD, Z boson, jet, differential, cross section}}

\maketitle

\section{Introduction}

\begin{figure}[!ht]
  \centering
  \includegraphics[width=0.5\textwidth]{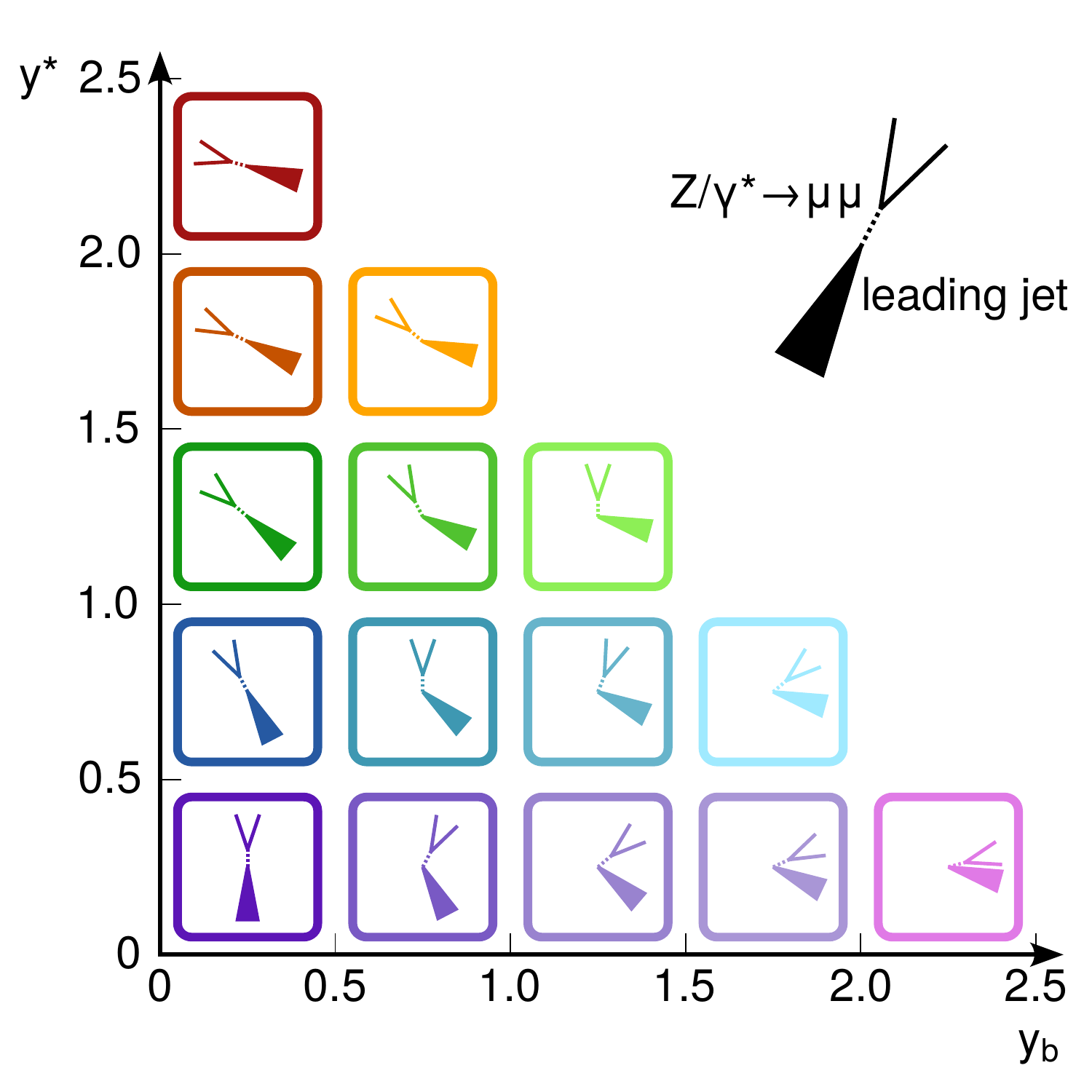}
  \caption{Illustration of the \zmmj event topologies and the phase space binning in the \ybys kinematic plane, where \ystar is half of the absolute rapidity
    separation between the \PZ boson and the leading jet, and \yboost is the boost of their center-of-mass system.
    For each bin, the \PZ boson and the leading jet are represented in the laboratory frame assuming a horizontal beam axis.
    To each \ybys bin an individual color is assigned that is used in further figures for better visual distinction.
  }
  \label{fig:phase_space}
\end{figure}

The production of a \PZ boson decaying into a pair of muons and associated with at least one jet (\zmmjx) is theoretically well
understood and has been measured precisely in proton-antiproton collisions at the Fermilab Tevatron~\cite{D0:2008lbc,D0:2009hdv,D0:2009rnw,CDF:2014hdx} and in proton-proton (\pp) collisions at the CERN LHC~\cite{CMS:2014bkk,CMS:2016php,CMS:2018mdf,ATLAS:2019bsa,ATLAS:2022nrp,CMS:2022ilp,ATLAS:2024xxl}.
This process can also be used to
constrain the partonic structure of the proton encoded in the form of parton distribution functions (PDFs) as performed, \eg, in
Refs.~\cite{Hou:2019efy,ATLAS:2021qnl,NNPDF:2021njg,Cridge:2023ozx}.  For this purpose, it is advantageous to subdivide the measurement phase
space into the following three dimensions: the \PZ boson transverse momentum, \ptz, half of the absolute rapidity separation between the \PZ boson and the leading
jet,
\begin{equation}
    \ystar = \frac{1}{2}\abs{\yz-\yjet},
\end{equation}
and the Lorentz boost of the \PZ boson and leading-jet system in the laboratory frame,
\begin{equation}
    \yboost = \frac{1}{2}\abs{\yz+\yjet}.
\end{equation}
This paper presents the first measurement using these observables, unfolded for detector effects in all three dimensions simultaneously.
The measurement is made possible by the large amount of data collected by the CMS experiment in 2016--2018.
While \ptz is representative of the energy scale of
the collision process, the variables \ystar and \yboost correlate with the scattering angle in the partonic center-of-mass frame of the $2 \to 2$ (partons${}\to\PZ{+}$jet) scattering process
and the imbalance in the initial-state parton momenta, respectively.
Assuming perfect balance between \ptz and the \pt of the leading jet, which is trivially valid at leading order (LO), the momentum fractions $x_{\pm}$ of the incoming partons can be expressed in terms of the final-state rapidities \yz and \yjet as
\begin{equation}
    x_{\pm} = \ptz \big( \exp(\pm \yz) + \exp(\pm \yjet) \big) / \sqrt{s},
\end{equation}
where $\sqrt{s}$ denotes the center-of-mass energy of the \pp collision.
Using the definitions of \ystar and \yboost, this expression can be rewritten as
\begin{equation}
    x_{\pm} = 2\ptz\,\cosh(\ystar)\exp(\pm \yboost) / \sqrt{s}.
\end{equation}
Hence, the cross section, triple-differential in these three observables, leads to constraints on the PDFs via comparisons with predictions at
next-to-next-to-LO (NNLO) in perturbative quantum chromodynamics (pQCD), as described in Section~\ref{sec:theory}.
Figure~\ref{fig:phase_space} illustrates the phase space and the binning in \ystar and \yboost used in this measurement.
Tabulated results are provided in the HEPData record for this analysis~\cite{HepData}.

This paper is organized as follows. A brief description of the CMS detector is given in Section~\ref{sec:detector}, followed by a description of the
analyzed data and simulation samples in Section~\ref{sec:samples}. The event reconstruction and selection are described in Section~\ref{sec:selection},
followed by the unfolding procedure in
Section~\ref{sec:unfolding}.  Experimental uncertainties are discussed in Section~\ref{sec:uncertainties} and the theory predictions in
Section~\ref{sec:theory}.  In Section~\ref{sec:results}, the results are compared with the predictions from Monte Carlo (MC) event generators and with the calculations
in pQCD at NNLO complemented with electroweak (EW) and nonperturbative (NP) corrections. Lastly, the paper is summarized in Section~\ref{sec:summary}.

\section{The CMS detector}
\label{sec:detector}

The central feature of the CMS apparatus is a superconducting solenoid of 6\unit{m} internal diameter, providing a magnetic field of 3.8\unit{T}.
Within the solenoid volume are a silicon pixel and strip tracker, a lead tungstate crystal electromagnetic calorimeter (ECAL), and a brass and
scintillator hadron calorimeter (HCAL), each composed of a barrel and two endcap sections.  Forward calorimeters extend the pseudorapidity ($\eta$) coverage
provided by the barrel and endcap detectors.  Muons are reconstructed in gas-ionization detectors embedded in the steel flux-return yoke outside the
solenoid.  More detailed descriptions of the CMS detector, together with a definition of the coordinate system and the relevant kinematic variables,
can be found in Refs.~\cite{CMS:2008xjf,CMS:PRF-21-001}.

Events of interest are selected using a two-tiered trigger system.
The first level, composed of custom hardware processors, uses information
from the calorimeters and muon detectors to select events at a rate of around 100\unit{kHz} within a fixed latency of 4\mus~\cite{CMS:2020cmk}.  The
second level, known as the high-level trigger, consists of a farm of processors running a version of the full event reconstruction software
optimized for fast processing, and reduces the event rate to a few kHz before data storage~\cite{CMS:2016ngn,CMS:2024aqx}.

\section{Data and simulated samples}
\label{sec:samples}

The \pp collision data used for this measurement were recorded by the CMS detector from 2016 to 2018 at a center-of-mass energy of
$\sqrt{s}=13\TeV$, corresponding to an integrated luminosity of 138\fbinv.  Events were selected with triggers requiring the presence of at
least one isolated muon with $\pt > 24\GeV$ in the 2016 and 2018 data-taking periods, and $\pt > 27\GeV$ in 2017.

The simulated signal sample for a pair of oppositely charged leptons in association with up to two jets is produced with \MGvATNLO 2.6.5~\cite{MadGraph,MadSpin,MCatNLO} at next-to-LO (NLO) in pQCD.
Samples with zero, one, and two jets in the final state are merged at NLO using the FxFx merging scheme~\cite{FxFx}.
The weights of the signal events are normalized to match the
cross section prediction of $6077\pm122\unit{pb}$ for the Drell--Yan process with a dilepton invariant mass above 50\GeV at NNLO accuracy in pQCD and NLO accuracy in
EW theory obtained with \FEWZ3.1.b2~\cite{Melnikov:2006kv,Gavin:2010az,Gavin:2012sy,Li:2012wna}.

Background contributions from top quark pair and single top quark production are simulated with
\POWHEG 2.0~\cite{Nason:2004rx,Frixione:2007vw,Alioli:2010xd,POWHEGWt,POWHEGtq,POWHEGhvq} at NLO accuracy.
The top quark pair event yields are normalized to match the prediction for the cross section of top-quark-antiquark pair
production at NNLO accuracy in pQCD of $831\,^{+40}_{-51}\unit{pb}$, provided by \TOPpp2.0~\cite{Toppp2} including the resummation of soft-gluon contributions up to next-to-next-to-leading logarithmic accuracy for a top quark mass of 172.5\GeV.
The events for the single top quark (antiquark) production in association with quarks are normalized to match the cross
sections of $\sigma(\PQt\PQq)=145.0\,^{+2.8}_{-1.9}\unit{pb}$ ($\sigma(\PAQt\PQq)=87.2\,^{+1.8}_{-1.5}\unit{pb}$) computed at NNLO in pQCD~\cite{STNNLOXS} using the same top quark mass.
The event yields for single top production associated with a \PW boson are normalized to a cross section of $71.7\pm3.9\unit{pb}$ obtained from calculations at NNLO
accuracy~\cite{NNLOST1,NNLOST2}.

Further background processes from diboson production are simulated with \POWHEG~\cite{Melia:2011tj,Nason:2013ydw} or \MGvATNLO at NLO accuracy.  Their cross sections are kept at NLO
accuracy as predicted by the respective MC generators.
Potential contamination from events with at least one nonprompt or misidentified muon is evaluated using a dedicated {\PW}+jets simulation sample at NLO for up to two jets using \MGvATNLO 2.6.5 and the FxFx merging scheme.
This contribution is found to be less than 0.01\% in every analysis bin and is, therefore, neglected.

Parton shower, underlying event, and hadronization are simulated using \PYTHIAviii~\cite{Sjostrand:2014zea} with the CP5
tune~\cite{CMS:2019csb} for all samples.
Additional \pp interactions from the same or nearby bunch crossings (pileup), simulated using minimum-bias events, are overlaid on the hard-scattering process with the multiplicity matching that in data.
In all simulated samples, the PDFs provided by PDF4LHC15~\cite{Butterworth2016} are used to describe the parton structure of the colliding protons.
This PDF set is a combination based on the CT14~\cite{Dulat2016}, MMHT14~\cite{HarlandLang2015}, and NNPDF3.0~\cite{Ball2015} PDF sets.
The detector response is simulated with the \GEANTfour package~\cite{Agostinelli:2002hh}.

\section{Event reconstruction and selection}
\label{sec:selection}

The particle-flow event reconstruction algorithm~\cite{CMS:2017yfk} aims to reconstruct and identify each individual particle in an event, using an
optimized combination of the information from all subdetectors.  In this process, the identification of the particle type, \ie, photon, electron, muon, charged
hadron, and neutral hadron, governs the determination of the particle direction and energy.  Photon candidates are identified as ECAL energy clusters
not linked to the extrapolation of any charged particle trajectory to the ECAL.  Electrons are identified as a primary charged particle track and
potentially multiple ECAL energy clusters corresponding to this track extrapolation to the ECAL, and to possible bremsstrahlung photons emitted along
the way through the tracker material.  Muons are identified as tracks in the central tracker consistent with either a track or several hits in the
muon system, and associated with calorimeter deposits compatible with the energy deposition of a muon on the corresponding trajectory.  Charged
hadrons are identified as charged particle tracks neither identified as electrons, nor as muons.  Finally, neutral hadrons are identified as HCAL
energy clusters not linked to any charged hadron trajectory, or as a combined ECAL and HCAL energy excess with respect to the expected charged hadron
energy deposit.

The energy of photons is obtained from the ECAL measurement.  The energy of electrons is determined from a combination of the track momentum at the
main interaction vertex, the corresponding ECAL cluster energy, and the energy sum of all bremsstrahlung photons attached to the track.  The energy of
muons is obtained from the corresponding track momentum.  The energy of charged hadrons is determined from a combination of the track momentum and the
corresponding ECAL and HCAL energies, corrected for the response of the calorimeters to hadronic showers.  Finally, the energy of neutral hadrons is
obtained from the corresponding corrected ECAL and HCAL energies.

For each event, hadronic jets are clustered from these reconstructed particles using the infrared and collinear safe anti-\kt
algorithm~\cite{Cacciari:2008gp, Cacciari:2011ma} with a distance parameter $R$ set to 0.4. Jet momentum is determined as the vectorial sum of all particle
momenta in the jet, and is found from simulation to be, on average, within 5--10\% of the true momentum over the entire \pt spectrum and detector
acceptance. Particles from pileup interactions can contribute additional tracks and calorimetric energy depositions clustered to the final jet momentum.
To mitigate this effect, charged particles identified as originating from pileup vertices are discarded,
and an offset correction is applied to correct for remaining contributions~\cite{CMS:PUMitigation}.
For the remaining jets, a classification based on boosted decision trees is performed to remove those originating from pileup.  Jet energy corrections are derived from simulation to
bring the measured response of jets to that of particle level jets on average. In situ measurements of the momentum balance in dijet,
photon+jet, {\PZ}+jet, and multijet events are used to account for any residual differences in the jet energy scale between
data and simulation~\cite{CMS:2016lmd}.  The jet energy resolution amounts typically to 15--20\% at 30\GeV, 10\% at 100\GeV, and 5\%
at 1\TeV.  Additional selection criteria are applied to each jet to remove jets dominated by anomalous
contributions from various subdetector components or reconstruction failures~\cite{CMS:PUMitigation}.

Muons are measured in the range $\abs{\eta} < 2.4$ using three technologies: drift tubes, cathode strip chambers, and resistive plate
chambers.  The single-muon trigger efficiency exceeds 90\% over the full $\eta$ range, and the efficiency to reconstruct and identify muons that occur
isolated from further event activity~\cite{CMS:2017yfk} is greater than 96\%.  To maximize the accuracy of the measurement, the muon four-momenta in
simulation are corrected to match the energy scale and resolution observed in data~\cite{RochesterCorr}.  Matching muons to tracks measured in the
silicon tracker results in a relative \pt resolution, for muons with \pt up to 100\GeV, of 1\% in the barrel and 3\% in the endcaps.
The \pt resolution in the barrel is better than 7\% for muons with \pt up to 1\TeV~\cite{CMS:2018rym}.

{\tolerance=800
In this analysis, events are selected by requiring at least two oppositely charged muons each with $\pt > 29\GeV$ and $\abs{\eta} < 2.4$. Candidate \PZ bosons are formed by combining two muons with opposite charge and a dilepton mass within $\pm$20\GeV of the \PZ
boson mass $\mZ = 91.1876\GeV$~\cite{ParticleDataGroup:2022pth}.  If multiple \PZ boson candidates can be reconstructed for the same event, the one closest to
\mZ is chosen.  The reconstructed \PZ boson is required to have $\ptz > 25\GeV$.  Additionally, at least one jet with
$\pt > 20\GeV$ and a rapidity $\abs{y} < 2.4$ is required. If a jet and a muon used in the reconstruction of the \PZ boson are
overlapping within a cone of $\DR = \sqrt{\smash[b]{(\Delta\phi)^2 + (\Delta\eta)^2}} < 0.4$, the jet is omitted.
Since jets are clustered from all visible final-state particles, this overlap removal avoids counting the two muons from the \PZ boson as part of any jet.
Only the jet with the highest \pt is considered to construct the observables of this analysis.
\par}

Event yields in simulation are scaled on an event-by-event basis to match the object selection and correction efficiencies measured in data.
Muon trigger, reconstruction, and identification efficiencies are derived with the ``tag-and-probe'' method~\cite{MuonReco7TeV}.  Efficiency corrections
for the pileup jet identification requirement are applied to simulated jets through scale factors that correct the efficiency difference between data and simulation.

An assignment of muon candidates to the wrong bunch crossing due to the limited time resolution of the muon systems leads to trigger inefficiencies.
This inefficiency, denoted as ``prefiring''~\cite{CMS:2021yvr,CMS:2020cmk}, varying between 0 and 3\%, affects the whole data set with muons and is most severe during 2016 data-taking.
Correction factors were computed from data and applied to the acceptance evaluated from simulation.

The selected events are binned according to their respective values in \ptz, \yboost, and \ystar to create a three-dimensional histogram with a total
of 264 bins for data, simulated signal, and simulated background.
The binning in \ystar and \yboost is done in steps of 0.5 ranging from 0 to 2.5, while the sum of \ystar and \yboost is restricted to be less than 3, as illustrated in Fig.~\ref{fig:phase_space}.
To account for different statistical precision, the binning in \ptz varies across the \ybys plane.  Three binning schemes are employed: generally the central binning (C) is used.  At the edge of the phase space,
where fewer events are expected, bins are merged (E) to ensure sufficient statistical precision.  Further merging (X) is
necessary in the highest \ystar bin to account for the even lower event yields. Also, the binning is truncated here at a maximum of 250\GeV. The binning
schemes are given in Table~\ref{tab:binning}.

\begin{table}[!ht]
  \centering%
  \topcaption{The bin borders in \ptz.  The central binning scheme (C) is used in the central rapidity regions.  At the edge of the \ybys plane, the edge
    binning scheme (E) is used, merging bins due to limited statistical precision.  An extra-wide binning scheme (X) is used for the highest \ystar bin. Due to the limited number of events, the X scheme truncates the binning at 250\GeV.}
  \renewcommand{\arraystretch}{1.2}
  \begin{tabular}{cp{0.75\textwidth}}
    Binning scheme & binning labels in \ybys plane{\huge\strut}\\
    \hline
                   & X \\
                   & E\quad{}E \\
    (cf.\ Fig.~\ref{fig:phase_space}) & C\quad{}C\quad{}E \\
                   & C\quad{}C\quad{}C\quad{}E \\
                   & C\quad{}C\quad{}C\quad{}C\quad{}E \\
    \hline
                   & bin edges in $\ptz\,/\GeV${\huge\strut}\\
    \hline
    Central (C) & 25, 30, 35, 40, 45, 50, 60, 70, 80, 90, 100, 110, 130, 150, 170, 190, 220, 250, 400, 1000\\
    Edge (E) & 25, 30, 35, 40, 45, 50, 60, 70, 80, 90, 100, 110, 130, 150, 170, 190, 250, 1000\\
    Extra (X) & 25, 30, 40, 50, 70, 90, 110, 150, 250 \\
  \end{tabular}\label{tab:binning}
\end{table}

\section{Unfolding}
\label{sec:unfolding}

Due to finite detector resolution and acceptance, the measured event yields per bin differ from the event yields at stable-particle level.
Migrations of events between bins due to resolution effects as well as detector inefficiencies need to be corrected to obtain the stable-particle level event yields.

\begin{figure}[!t]
  \centering%
  \includegraphics[width=\textwidth]{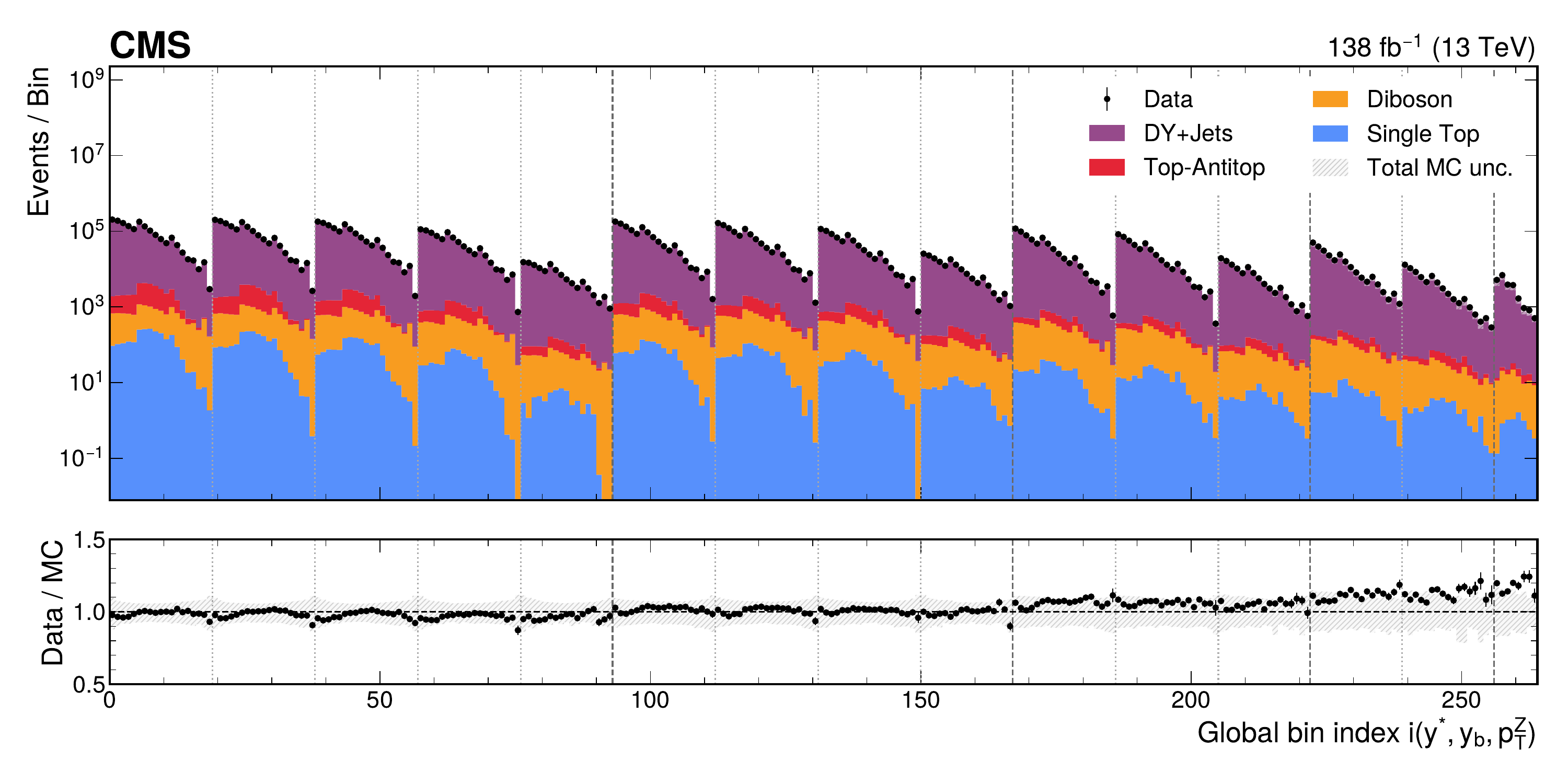} \\
  \includegraphics[width=\textwidth]{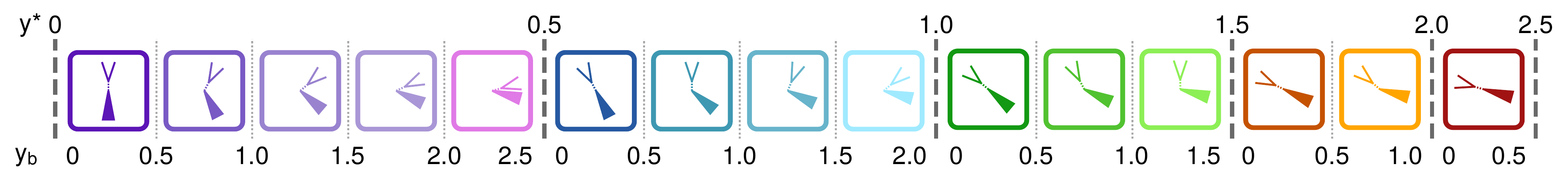}
  \caption{Comparison of data and simulation at the reconstruction level for the whole unrolled three-dimensional phase space in \ystar, \yboost, and
    \ptz. The nonnormalized event counts are shown per bin, so changes in binning width induce visible structures.  The lower panel shows the ratio of data to simulation. The error bars on the data points represent the statistical uncertainties, while
    the gray band indicates the total uncertainty in the simulation. The gray dotted and black dashed vertical lines indicate the bin
    borders in \yboost and \ystar, respectively. The unrolling in \ybys is illustrated below the plot. Each bin in \ybys contains the respective series of \ptz bins.} \label{fig:data_mc_unrolled}
\end{figure}

The signal and background event yields are obtained from simulation, as described in Section~\ref{sec:samples}, and compared with the event yields in data.
The agreement between data and MC simulation at the reconstruction level is shown in Fig.~\ref{fig:data_mc_unrolled} for the whole unrolled three-dimensional phase space.
In order to obtain a one-dimensional representation of the triple-differential phase space, the bins are unrolled in the order of increasing \ystar,
\yboost, and \ptz.
The 264 unrolled bins are enumerated by a global bin index $i$.
Overall, a good agreement between data and simulation, deteriorating with increasing \ystar, is observed.
This behavior is expected, due to the large NLO-to-LO correction factors (``$K$ factors'', \ie
ratios of higher-order to LO cross sections) at high \ystar, essentially reducing the NLO MC predictions to merely LO accuracy
in these regions, as explained in Section~\ref{sec:theory}.
The uncertainty in the simulation includes statistical and all systematic effects, as described in Section~\ref{sec:uncertainties}.
Additionally, theoretical uncertainties in the signal prediction are included, determined in the same way as the corresponding background prediction uncertainties, also described in Section~\ref{sec:uncertainties}.

\begin{figure}[!ht]
  \centering%
  \includegraphics[width=\textwidth]{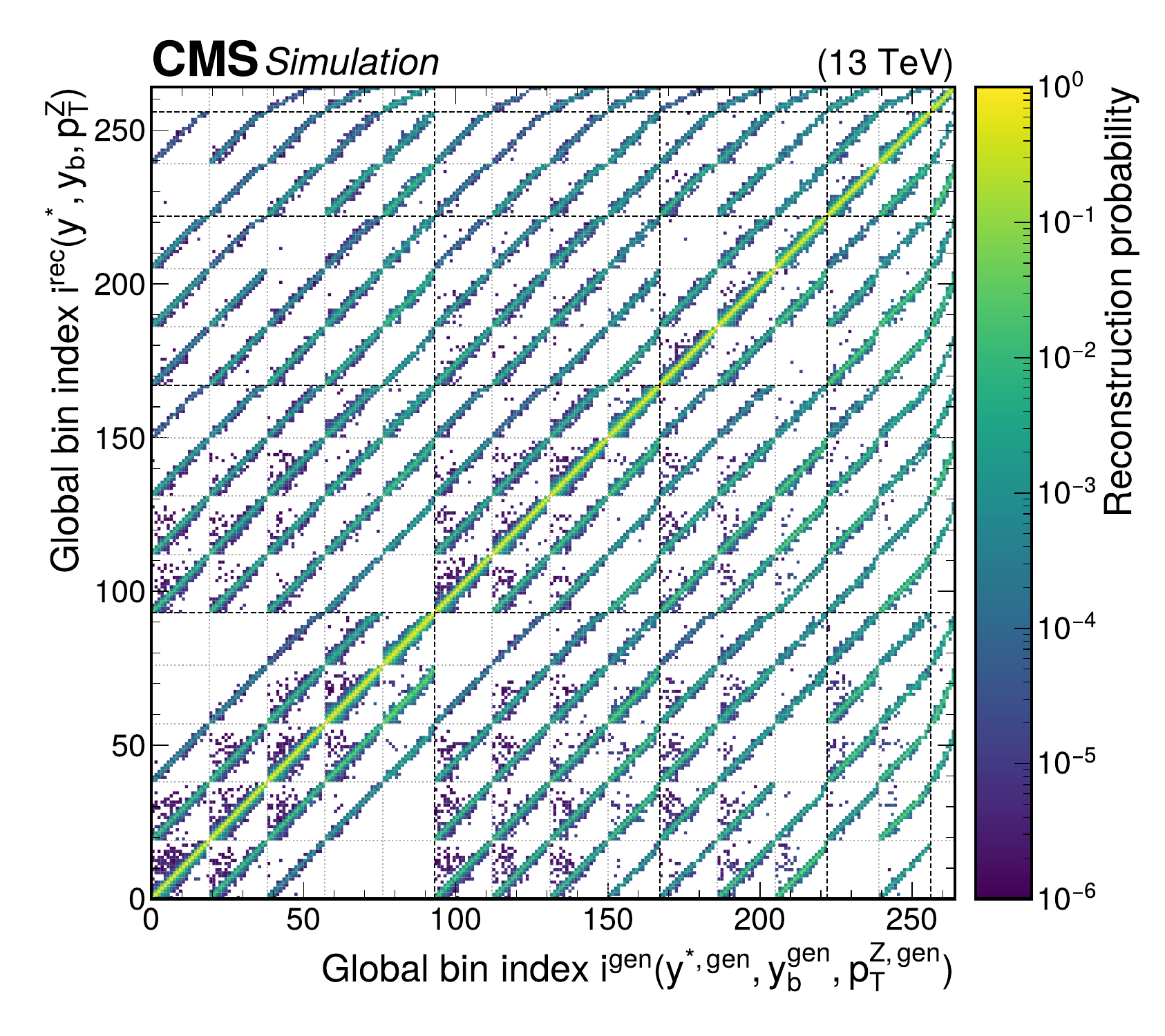}
  \caption{Detector response matrix for the unrolled three-dimensional phase space of \ystar, \yboost and \ptz.  The $x$- and $y$-axes correspond to the
    global bin indices $i^{\text{gen}}$ at the particle and $i^{\text{rec}}$ at the reconstruction level, respectively.  The colored boxes indicate the
    probability to reconstruct an event in the reconstruction bin number $i^{\text{rec}}$, given it had been generated in the generation bin
    number $i^{\text{gen}}$. Everything not on the principal diagonal corresponds to event migrations.}\label{fig:response}
\end{figure}

The detector response is estimated from simulation.
A two-dimensional response matrix, with each dimension containing the same 264 unrolled \ystar, \yboost, and \ptz bins, is filled on an event-by-event basis by evaluating the
observables at the particle and reconstruction levels.  The resulting response matrix with
indexed bins is shown in Fig.~\ref{fig:response}.
Because of the one-dimensional representation of the phase space, neighboring bins in \ybys are not always next to each other.
At the edge of the matrix in the high-\ystar bin, the reduced extra binning scheme in \ptz (shown in Table~\ref{tab:binning}) is employed.
This results in a visually different slope for the diagonals of this bin with respect to migrations in comparison with all other \ybys bins.

Before the unfolding, the background contributions estimated from simulation are subtracted.  Additionally, the relative fraction of simulated signal
events passing the selection criteria at detector level but not at stable-particle level is estimated from simulation.  This arises from events moving into
the selected phase space, \eg, due to identifying pileup jets as jets from the hard process or other smearing effects at the edge of the phase space
during reconstruction.  The estimated rate decreases with \ptz from about 25\% to less than 1\% and is subtracted from the measured event yields in data before unfolding.

The unfolding is performed using matrix inversion provided by the \TUnfold package~\cite{Schmitt:2012kp} based on the response matrix derived from
simulation.
The condition number of the response matrix, defined as the ratio of its largest to smallest singular value and, therefore, a measure of sensitivity to numerical fluctuations, is 2.8, which is sufficiently small to allow for a stable unfolding using matrix inversion without regularization.
Additionally, acceptance effects from the event selection, such as trigger, reconstruction, and identification efficiencies, are estimated from simulation and corrected for during the unfolding. The associated uncertainties are discussed in Section~\ref{sec:uncertainties}.

The stability of the unfolding is verified with alternative simulation models.
First, two simulations of \PZ boson production in association with up to four jets at LO accuracy in pQCD with \MADGRAPH using MLM merging~\cite{Alwall:2007fs} are investigated.
One sample is interfaced to \PYTHIAviii, and the other to \HERWIG 7.1.4~\cite{Bellm:2015jjp} using the CH3~\cite{CMS:2020dqt} tune.
Neither sample describes the normalization or the shape of the data within uncertainty.
Therefore, the unfolding is repeated using a simulation with the
\minnlops~\cite{Monni:2019whf,Monni:2020nks} formalism within the \POWHEG framework interfaced with \PYTHIAviii and \PHOTOS 3.64~\cite{Barberio:1993qi}.
This prediction is NNLO accurate in the zero-jet phase space of the Drell--Yan production, while being NLO accurate in the one-jet phase
  space, which is relevant in this analysis.
For this sample, the description of the data is of similar quality to that of the nominal \MGvATNLO sample.
After unfolding, the two unfolded results differ by 1\% up to 4\% at most in the central bins with $\ystar<0.5$ and $\yboost<1.5$, and for $\ystar<1.0$ and $\yboost<1.0$.
At larger \ystar and \yboost, both samples yield unfolded results compatible within the statistical uncertainties in the simulation.
Since neither of the two NLO signal simulations is preferred a priori, they are treated as two equally probable models of the detector response.
The central value of the measurement is therefore taken as the average of the two unfolded results, and half of their difference is assigned as a
systematic unfolding uncertainty, further detailed in
Section~\ref{sec:uncertainties}.

\section{Experimental uncertainties}
\label{sec:uncertainties}

\begin{figure}[!t]
  \centering%
  \includegraphics[width=0.5\textwidth]{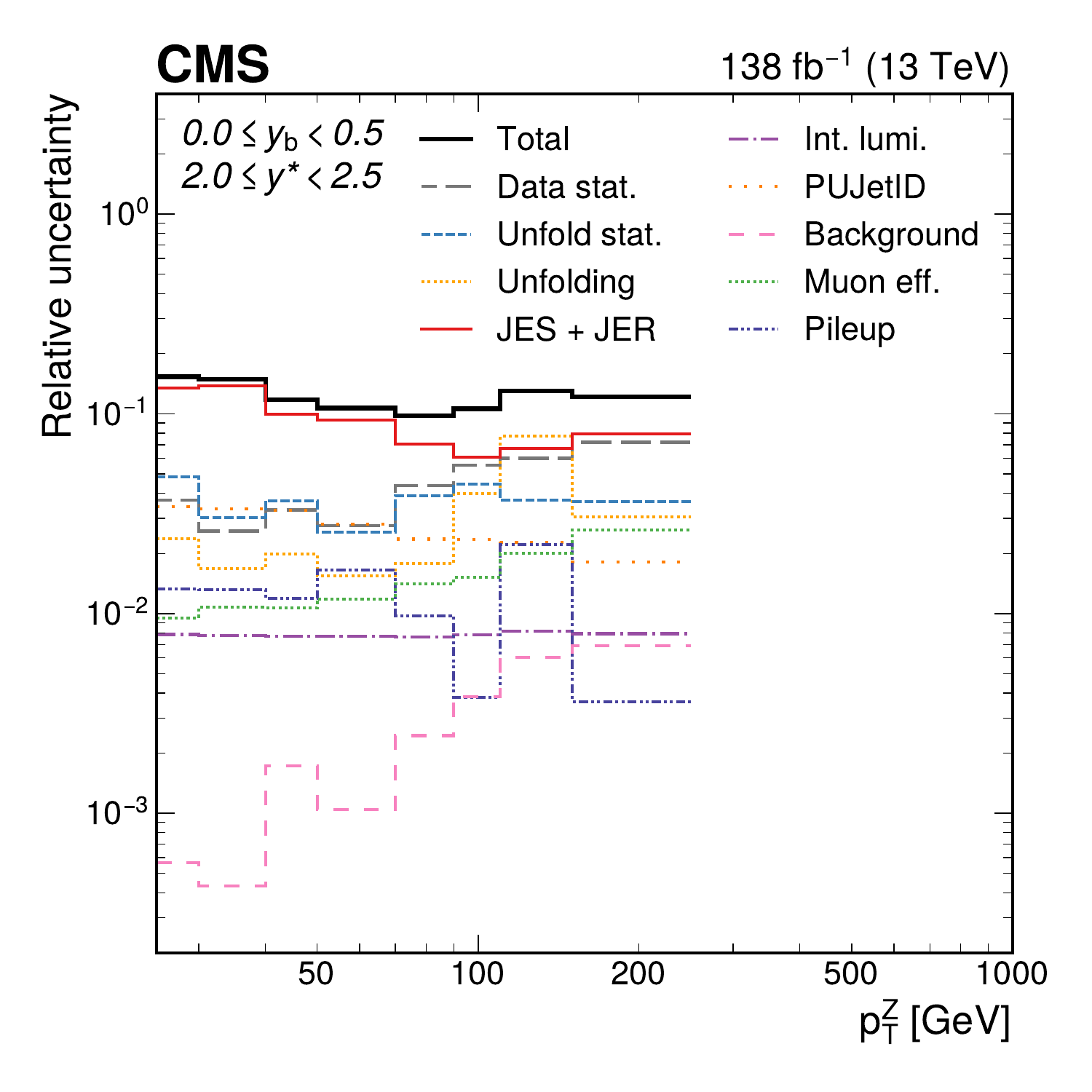}%
  \includegraphics[width=0.5\textwidth]{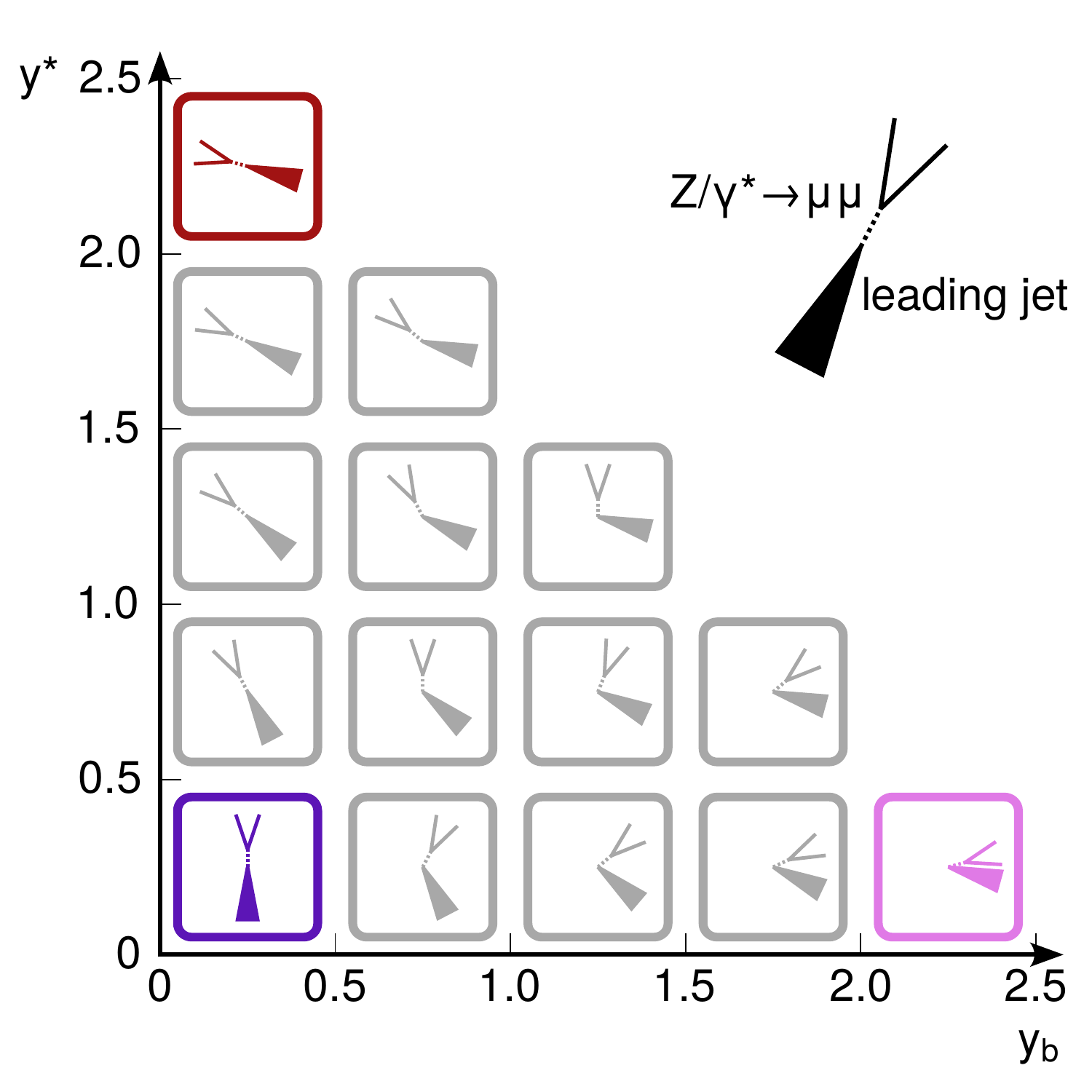}\\
  \includegraphics[width=0.5\textwidth]{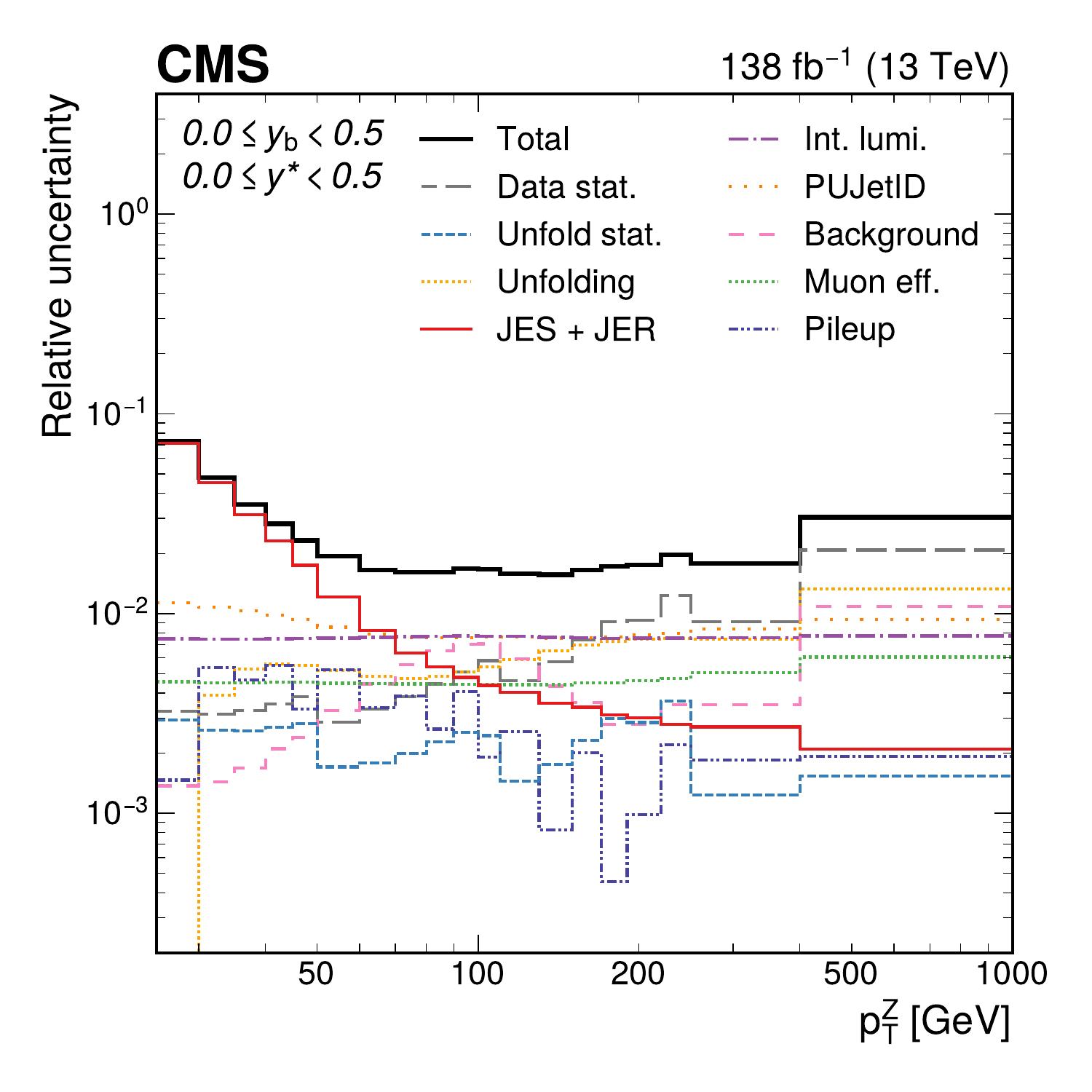}%
  \includegraphics[width=0.5\textwidth]{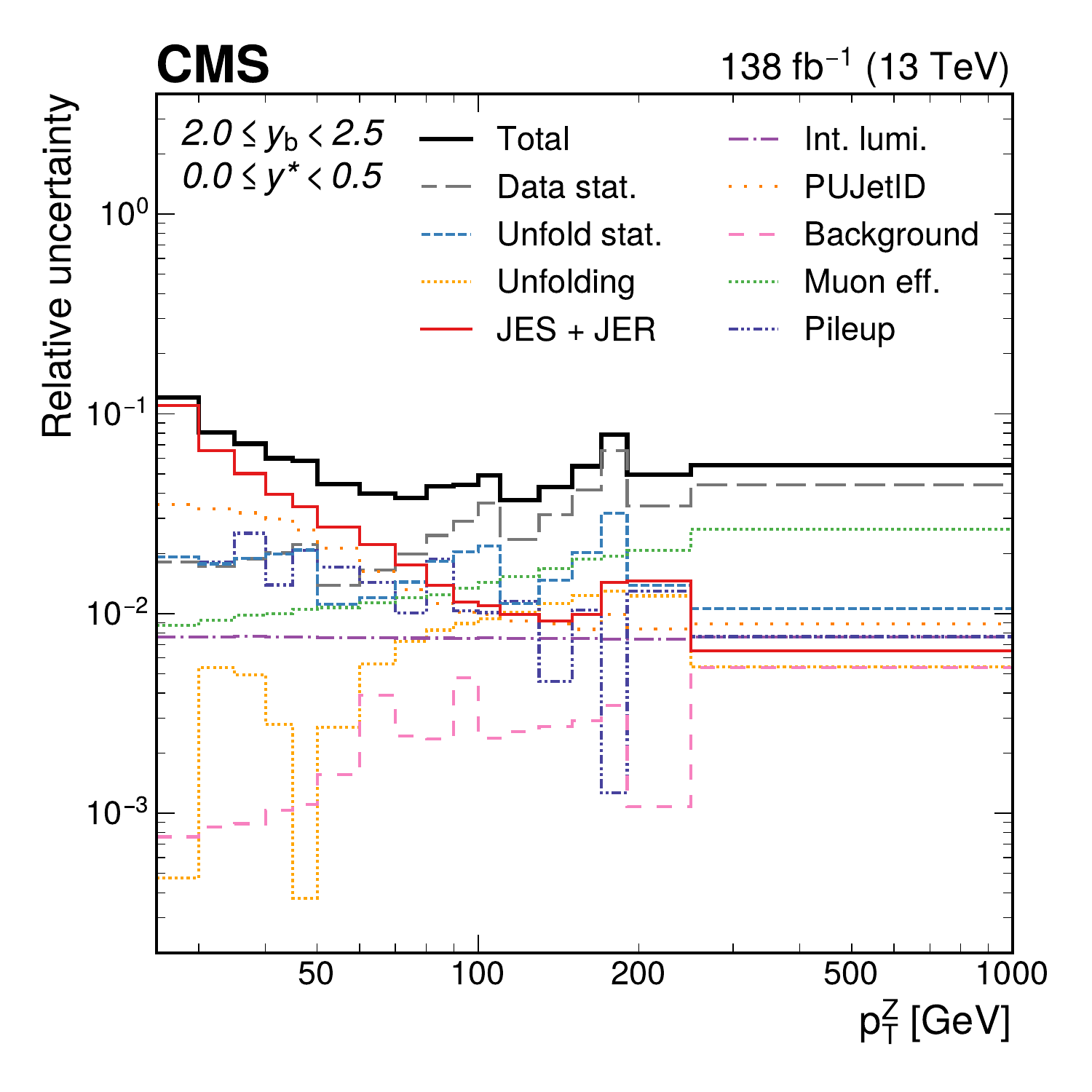}
  \caption{%
    Systematic uncertainties in the unfolded \zmmjx cross section as functions of \ptz for three phase space intervals in the \ybys kinematic plane.
    An illustration of the presented \ybys bins is shown in the top right.
    The upper left plot shows the uncertainties for a forward-backward topology (large \ystar) of the $2\to 2$ scattering process.
    In the lower row the uncertainties are presented for a central back-to-back topology, where the \PZ boson and the balancing jet are scattered in
    opposite directions perpendicularly to the beam (left) and a boosted topology (large \yboost, right) of the $2\to 2$ scattering process.
    Labels for the various uncertainty components are explained in the text.
    For a better visual comparison, the range in \ptz is kept the same for all plots.
  }\label{fig:uncertainties}
\end{figure}

The uncertainties in the unfolded results are given in Fig.~\ref{fig:uncertainties}, which includes the decomposition into several sources.
The statistical uncertainties in the data, labeled ``Data stat.'', are calculated from the event counts in each bin assuming a Poisson distribution.  Overall, the limited
number of events in the data leads to statistical uncertainties of 0.5\% in the central rapidity bins and up to 6.5\% in the high-\ystar and
-\ptz bins.  Additionally, the statistical uncertainty of the response matrix and of the acceptance are considered.  The
statistical uncertainty in the MC sample including correlations between bins is derived by \TUnfold and propagated to the final results.  It
ranges from 0.25\% in the central rapidity region and low-\ptz bins up to 4.5\% in the highest \ystar and \ptz bins.
Furthermore, the statistical uncertainty in the out-of-acceptance contribution (events passing the selection criteria at detector but not at stable-particle level) is considered, which is well below 1\% in the central and high-\yboost rapidity regions and reaches up to 3\% in the high-\ystar and low-\ptz bins.
The statistical uncertainties arising from the response matrix and the out-of-acceptance contribution are uncorrelated and added in quadrature, labeled ``Unfold stat.'' in Fig.~\ref{fig:uncertainties}.

Uncertainties from systematic effects are taken into account during the unfolding procedure.  For each systematic uncertainty, the unfolding procedure
is repeated with a modified response matrix, acceptance, and out-of-acceptance contribution estimation.  Uncertainties are then constructed using the deviations
from the nominal unfolded cross section to the results obtained for each $\pm$1 standard deviation shift for each uncertainty source.
This includes uncertainties in the unfolding model, the jet energy scale and resolution, the integrated luminosity, the pileup and background estimation, the pileup jet identification, and in the muon trigger, reconstruction, identification, and prefiring corrections. In this way, the procedure also propagates effects induced by varying the reconstructed observables within their uncertainties, including cases where the leading jet is exchanged with a subleading jet.

The choice of the underlying theory model from which the response matrix, acceptance, and out-of-acceptance contribution are derived affects the unfolded result.
As described in Section~\ref{sec:unfolding}, the complete unfolding procedure is therefore
repeated using the \POWHEG{}\,\minnlops + \PYTHIAviii sample in place of the nominal \MGvATNLO sample.
To avoid statistical fluctuations in the unfolding uncertainty---the statistical component is covered by the ``Unfold stat.'' term---the relative difference between the two results is evaluated from smoothed versions of the unfolded distributions.
Half of this relative difference is assigned as a systematic uncertainty for the choice of the underlying theory model and is labeled as ``Unfolding'' in Fig.~\ref{fig:uncertainties}.
The resulting uncertainty amounts to about 0.5\% in most of the phase space and stays below 1.4\% in the
central \ybys region.  It grows to 3\% up to at most 8\% in the highest-\ystar bins, where it becomes comparable to the statistical uncertainty.

The calibration of the jet energy scale (JES) induces uncertainties up to 14\% for low \ptz in the outer rapidity bins, decreasing with higher
\ptz.  In the central rapidity region, the maximum JES uncertainty is less than 8\% for the low-\ptz region and decreases to less than 0.5\%
in the high-\ptz region. Uncertainties in the jet energy resolution (JER) affect the cross section by less than 0.5\% across all bins.
Both uncertainties are uncorrelated and added in quadrature, labeled ``JES+JER'' in Fig.~\ref{fig:uncertainties}.

The integrated luminosities for the 2016, 2017, and 2018 data-taking years have 0.82--1.2\% individual
uncertainties~\cite{CMS-LUM-17-003,CMS-PAS-LUM-20-001}, while the overall uncertainty for the combined period is 0.73\%.
The uncertainties in the individual data-taking years are taken into account including their year-to-year correlations when creating the combined response
matrix for the unfolding and are propagated to the final results.
The final uncertainty in the integrated luminosity is labeled ``Int.\ lumi.'' in Fig.~\ref{fig:uncertainties}.

The number of pileup interactions per event in simulation is varied by changing the total inelastic cross section of 69.2\unit{mb} by $\pm$4.6\%.
The resulting uncertainty in the final result, labeled ``Pileup'' in Fig.~\ref{fig:uncertainties}, is below 1\% in the central rapidity region and
goes up to 2\% in the outer rapidity regions.

The uncertainty in the simulated background event yields is composed of multiple sources.
To estimate the effect of missing higher orders in pQCD, the renormalization scale \mur and
factorization scale \muf in the MC generators are varied independently by factors of 0.5 and 2 relative to their nominal value \mun. The
uncertainty is estimated from the maximum variation of the resulting predictions from the nominal, excluding the combinations $(\mur/\mun, \muf/\mun) = (0.5, 2.0)$ and
$(2.0, 0.5)$, where the ratio of the two scales differs by a factor of four.  The PDF uncertainty is evaluated following the prescription for
PDF4LHC15~\cite{Butterworth2016} using a set of symmetric Hessian eigenvectors~\cite{Gao2014,Carrazza2015,Watt2012} and including variations of
the strong coupling constant \asmz by $\pm$0.0015 around its central value of 0.118.  In addition, uncertainties related to the modeling of initial- and final-state radiation are assessed by varying the corresponding scales
by a factor of 0.5 and 2.
Since the background only contributes about 1\% of the total event yield, and it does not exceed 6\% in any bin, the uncertainties in the background simulation result in a minor impact on the final result.
The contributions from nonresonant backgrounds, namely top quark pair and single top quark production, are assigned an
additional 10\% normalization uncertainty to cover potential mismodeling effects, based on comparisons with $\Pe\PGm$ data performed in Ref.~\cite{CMS:2022ilp}.
The size of this additional uncertainty in the final results is less than 0.1\% due to the low overall background and is negligible compared with other uncertainty sources.
Furthermore, the statistical uncertainty from the limited number of simulated background events is taken into account.
Each systematic uncertainty source for the total background estimation is evaluated simultaneously for all MC samples.
Subsequently, all uncertainty components are added in quadrature, labeled ``Background'' in Fig.~\ref{fig:uncertainties}.

The uncertainty caused by the pileup jet mitigation, labeled ``PUJetID'' in Fig.~\ref{fig:uncertainties}, reaches up to 3\% in the high-\ystar bins for low \ptz, but remains around 1\% for low
\ptz in the rest of the phase space.

Corrections applied to the simulation samples to account for the trigger, reconstruction, and identification efficiencies of the muons, as well as
uncertainties in the trigger prefiring, are below 0.5\% in most bins, rising up to 2\% in the outermost \ystar and \ptz bin.
These uncertainties are uncorrelated and added in quadrature, labeled ``Muon eff.'' in Fig.~\ref{fig:uncertainties}.

{\tolerance=800
The statistical uncertainty and different systematic sources are uncorrelated and added in quadrature to obtain the total uncertainty shown in Fig.~\ref{fig:uncertainties}. At low \ptz, the main contribution stems from the JES uncertainty, while at high \ptz the statistical uncertainty becomes dominant.
In the central \ybys bins the highest accuracy is achieved with a precision of 1.5\%.
Here, multiple sources of similar magnitude contribute to the total uncertainty, including the integrated luminosity, pileup jet identification, the unfolding model, and background estimation.
\par}

\section{Theory predictions}
\label{sec:theory}

\begin{figure}[!tp]
  \centering%
  \includegraphics[width=0.5\textwidth]{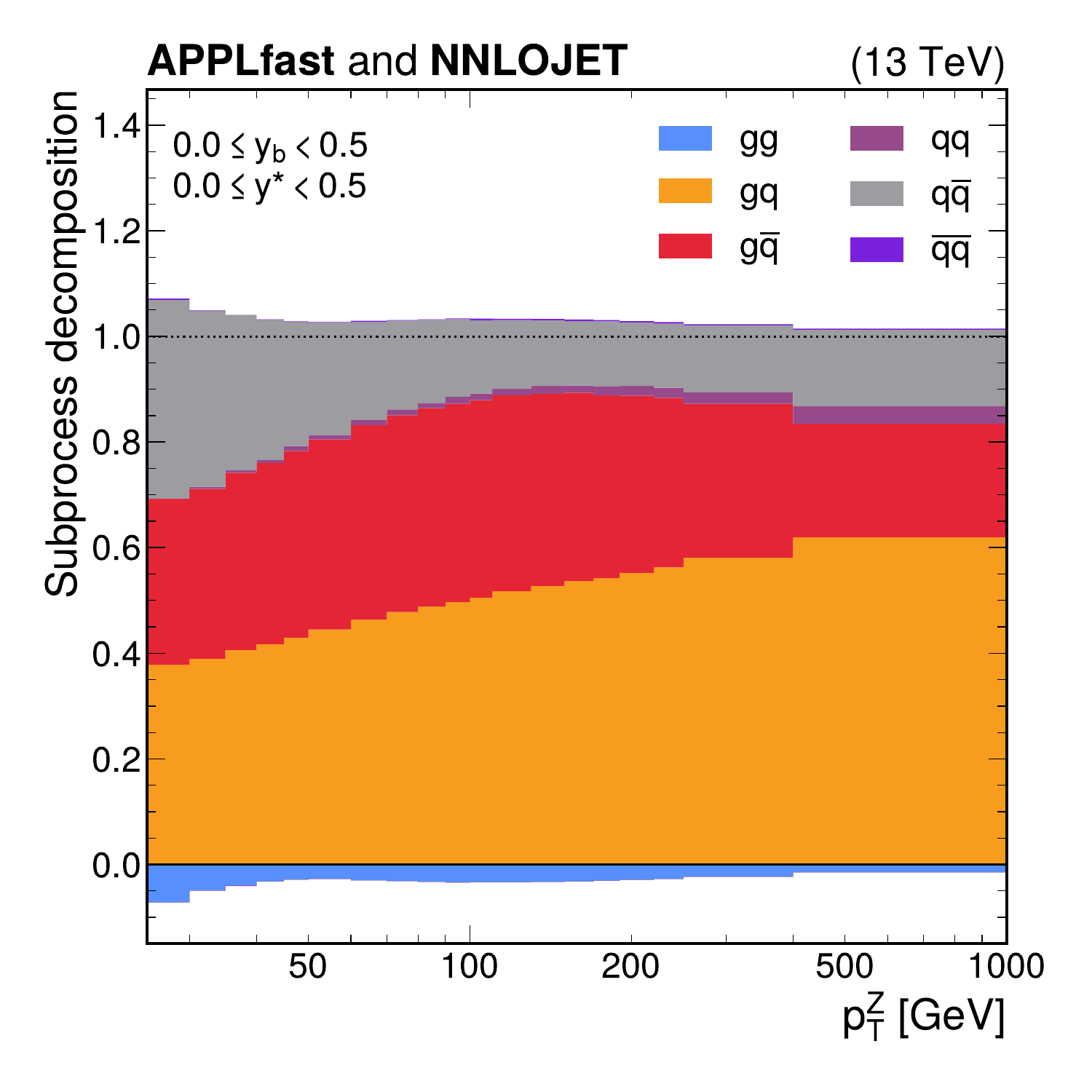}%
  \includegraphics[width=0.5\textwidth]{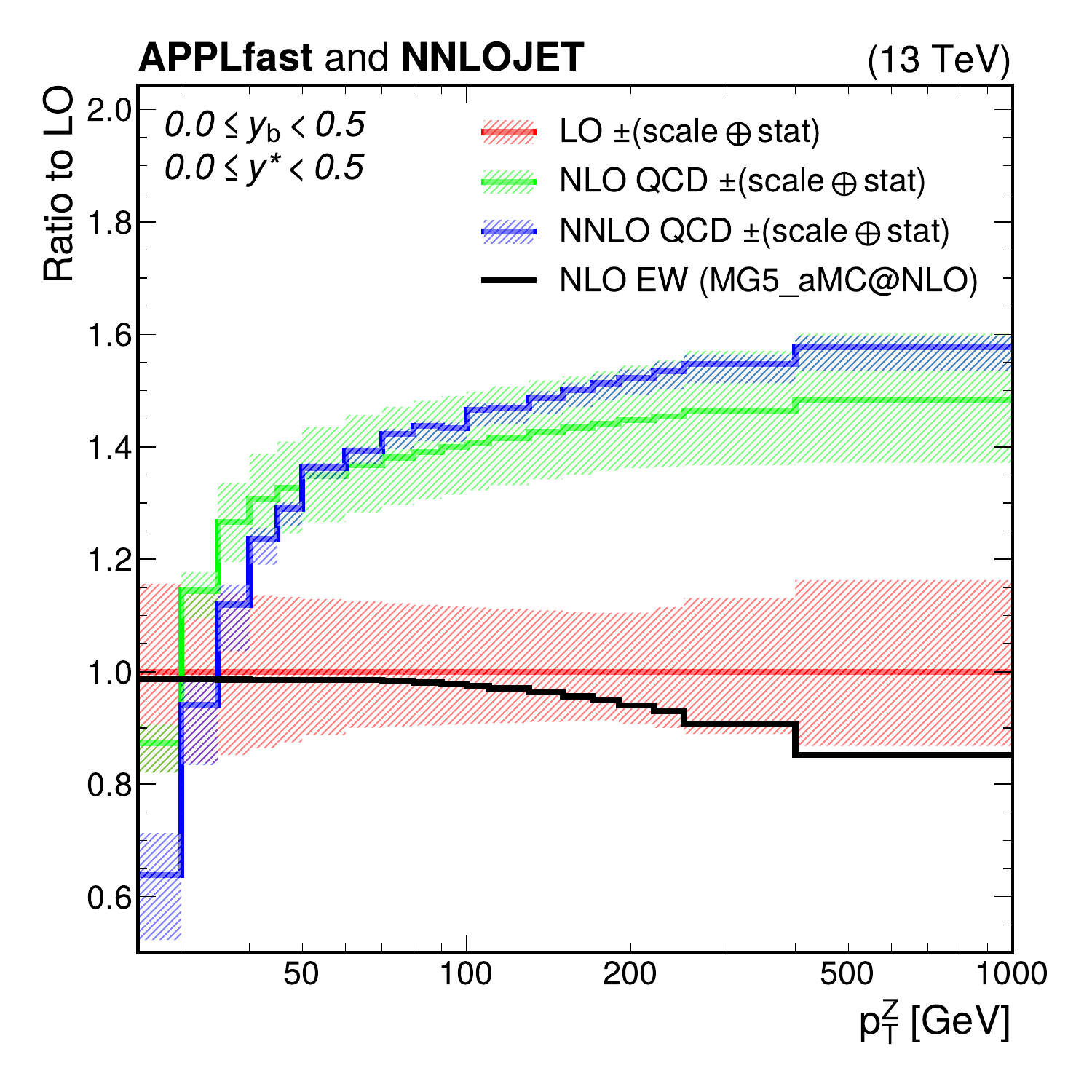}\\
  \includegraphics[width=0.5\textwidth]{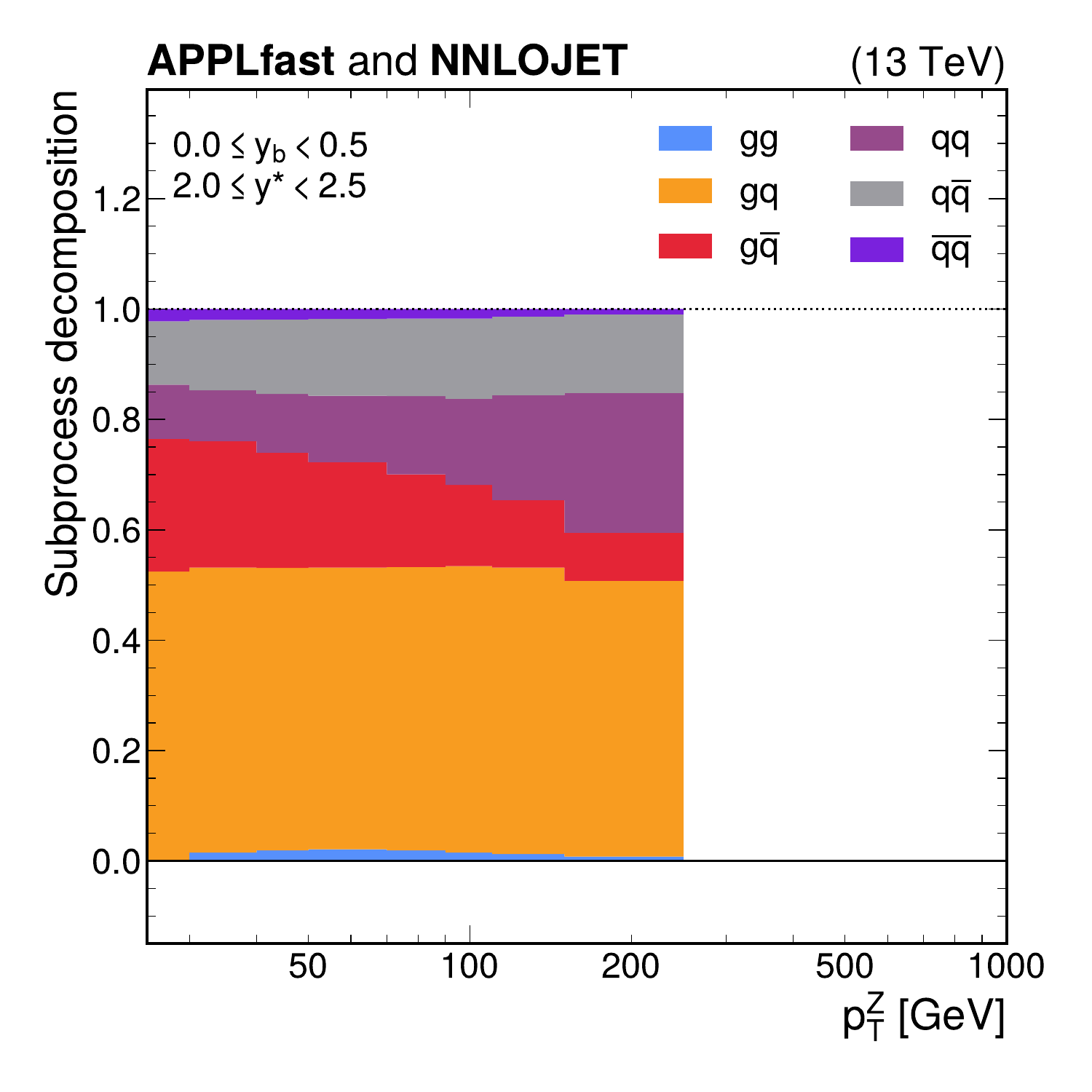}%
  \includegraphics[width=0.5\textwidth]{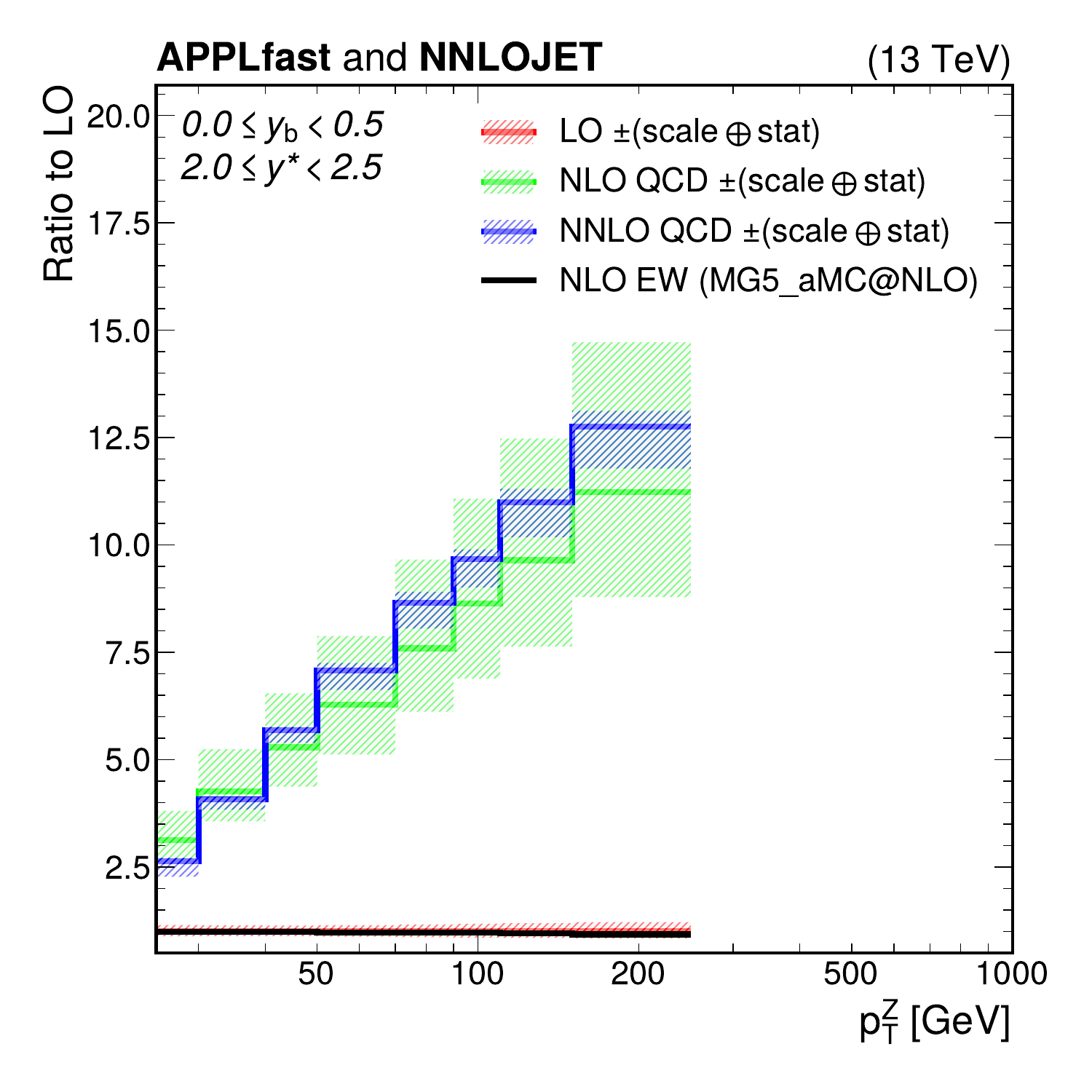}
  \caption{%
    Subprocess decomposition at NNLO (left) and ratios ($K$ factors, right) of the cross-section predictions from \NNLOJet at NLO (green) and NNLO (blue) in pQCD to LO (red) both as a function of \ptz for two phase space intervals in the \ybys kinematic plane.
    Additionally, the NLO EW correction compared with LO obtained from \MGvATNLO is shown in black.
    The upper row corresponds to a central back-to-back topology, where the \PZ boson and the balancing jet are scattered in opposite directions perpendicularly to the beam.
    The lower row presents the results for a forward-backward topology with large rapidity separation \ystar between the \PZ boson and the balancing jet.
    All predictions are based on the PDF4LHC21 PDF set at NNLO\@.
    The colored areas on the left correspond to the relative cross section contributions from initial state parton-parton combinations as given in the legend.
    Interference terms appearing in the calculations beyond tree level can lead to negative parton-parton cross sections.
    The colored bands on the right indicate the uncertainty in a prediction
    based on scale variations, as further described in the text.
    For a better visual comparison, the range in \ptz is kept the same for all plots.
  }
  \label{fig:pqcd}
\end{figure}

Including predictions for \zmmj production up to NNLO in pQCD, which have first been calculated in
Refs.~\cite{Boughezal:2015ded,Ridder:2015dxa,Ridder:2016nkl}, the cross section can be written as
\begin{equation}
  \sigQCDNNLO = \sigLO + \delta\sigQCDNLO + \delta\sigQCDNNLO,
\end{equation}
where the LO prediction \sigLO corresponds to the matrix element of order $\ord(\as^1\alpha^2)$.
Here, \as and the electromagnetic coupling constant $\alpha$ are used for the power counting of the expansion in perturbation theory.
The additional pQCD terms at NLO
($\delta\sigQCDNLO$) and NNLO ($\delta\sigQCDNNLO$) each increment the power in \as by one, to
$\ord(\as^2\alpha^2)$ and $\ord(\as^3\alpha^2)$, respectively. Here, they are derived with the \NNLOJet~\cite{Gehrmann:2018szu} program interfaced
via \APPLfast~\cite{Britzger:2022lbf} to the \FastNLO~\cite{Britzger:2012bs} package for the preparation of interpolation grids. Such grids enable
fast reevaluations of the NNLO cross sections for varying PDFs, \asmz, or scale factors of \mur and \muf, as described in
Section~\ref{sec:uncertainties}.  Moreover, as illustrated in Fig.~\ref{fig:pqcd} (left column), the relative parton-parton subprocess decomposition can be
obtained.

Figure~\ref{fig:pqcd} (upper left) presents the subprocess decomposition at NNLO for a topology with the \PZ boson and the balancing jet
being scattered in opposite directions perpendicular to the beam, requiring roughly equal parton momentum fractions: $x_+ \approx x_-$.  The
dominating parton-parton subprocesses in this phase space region are gluon-(anti-)quark and quark-antiquark scattering (red, orange, and gray),
possible already at LO\@.
The forward-backward topology shown in Fig.~\ref{fig:pqcd} (lower left) has a large rapidity separation \ystar between the \PZ boson and the balancing
jet, approaches the kinematic limit at large \ptz, and similarly demands equal, but much larger parton momentum fractions $x_+$ and $x_-$. This
configuration is strongly suppressed at LO because the gluon and the antiquark momentum distributions favor small values of the fractional momentum
$x$. The substantially contributing quark-quark subprocess (light purple) starts being possible only at NLO\@. In contrast, the boosted regime (not shown)
for large \yboost requires $x_+/x_-$ far away from unity and is not ``PDF suppressed''. The subprocess decomposition and the reach in \ptz in this case
resemble the one for the central back-to-back topology.

Figure~\ref{fig:pqcd} (right column) presents, for the same two topologies, the ratios (or $K$ factors) of the higher-order calculations with respect to the LO
one including the respective uncertainty estimates shown as bands that are based on the \mur and \muf scale variations described in
Section~\ref{sec:uncertainties}. Although the NLO prediction lies outside the LO uncertainty band for the central back-to-back topology, a very
reasonable perturbative convergence is exhibited in this region with small $K$ factors and the NNLO result within the NLO uncertainty band except for
the lowest \ptz bins, which are close to acceptance thresholds.
In contrast, because of the PDF-suppressed LO processes, the forward-backward region suffers from very large NLO $K$ factors. Nevertheless, the NNLO result
again is only a small correction within the uncertainty of the NLO one.

Similarly conceived triple-differential dijet measurements have been published in Refs.~\cite{CMS:2017jfq,CMS:2023fix}.  In contrast to the relative
parton-parton subprocess decomposition obtained here for \zmmj, the one for dijet production as shown, \eg, in Fig.~2 of Ref.~\cite{CMS:2017jfq}, looks very
different.  While the gluon-gluon channel contributes significantly to dijet production at small transverse momentum, and the quark-quark channel at
large transverse momentum, both come into play only at NLO for the \zmmj process.  The presented triple-differential measurement of \zmmj therefore
provides valuable complementary information for constraining the partonic structure of the proton.

At NNLO accuracy in pQCD additional EW effects become sizable compared with the QCD corrections and must be considered, in particular for large \ptz.
The cross section for NLO EW predictions including contributions up to order $\ord(\as\alpha^3)$ can be written as
\begin{equation}
  \sigEWNLO = \sigLO + \delta\sigEWNLO.
\end{equation}
{\tolerance=800
Using \MGvATNLO~\cite{Frederix:2018nkq} version 3.7.0, NLO EW corrections for \zmmj production are calculated and shown as a ratio to the LO prediction in Fig.~\ref{fig:pqcd} (right column).
The NLO EW corrections reduce the cross section by up to 15\% at high \ptz, which is of similar magnitude as the NNLO to NLO pQCD correction.
The fixed-order calculations with \MGvATNLO use the PDF4LHC21 PDF set~\cite{PDF4LHCWorkingGroup:2022cjn} and have been cross-checked against the results from \NNLOJet at LO, yielding the same values within statistical uncertainties.
Additional contributions from photon initial states were assessed using the LUXqed17 plus PDF4LHC15 PDF set~\cite{Manohar:2017eqh} and were found to be negligible.
\par}

Assuming the EW corrections to factorize with respect to the higher-order pQCD corrections, their multiplicative combination can be defined as
\begin{equation}
  \sigma_{\text{QCD}\times\text{EW}}^{\text{NNLO,NLO}} = \sigQCDNNLO K_{\text{EW}},
  \quad \text{with} \quad
  K_{\text{EW}} = 1+\frac{\delta\sigEWNLO}{\sigLO}.
\end{equation}
Lacking mixed correction terms of order $\ord(\as^2\alpha^3)$, the multiplicative approach might be insufficient to describe the full higher-order EW effects in regions of large QCD $K$ factors.
Therefore, half the difference between the multiplicative and additive combination of EW and QCD corrections is used as an additional uncertainty,
where the additive combination of the NNLO pQCD and NLO EW corrections is defined as
\begin{equation}
  \sigma_{\text{QCD}+\text{EW}}^{\text{NNLO,NLO}} =
  \sigLO + \delta\sigQCDNLO +
  \delta\sigQCDNNLO + \delta\sigEWNLO.
\end{equation}
This results in an uncertainty of up to 4\% to the cross section predictions at high \ptz.

\begin{figure}[!tp]
  \centering%
  \includegraphics[width=0.5\textwidth]{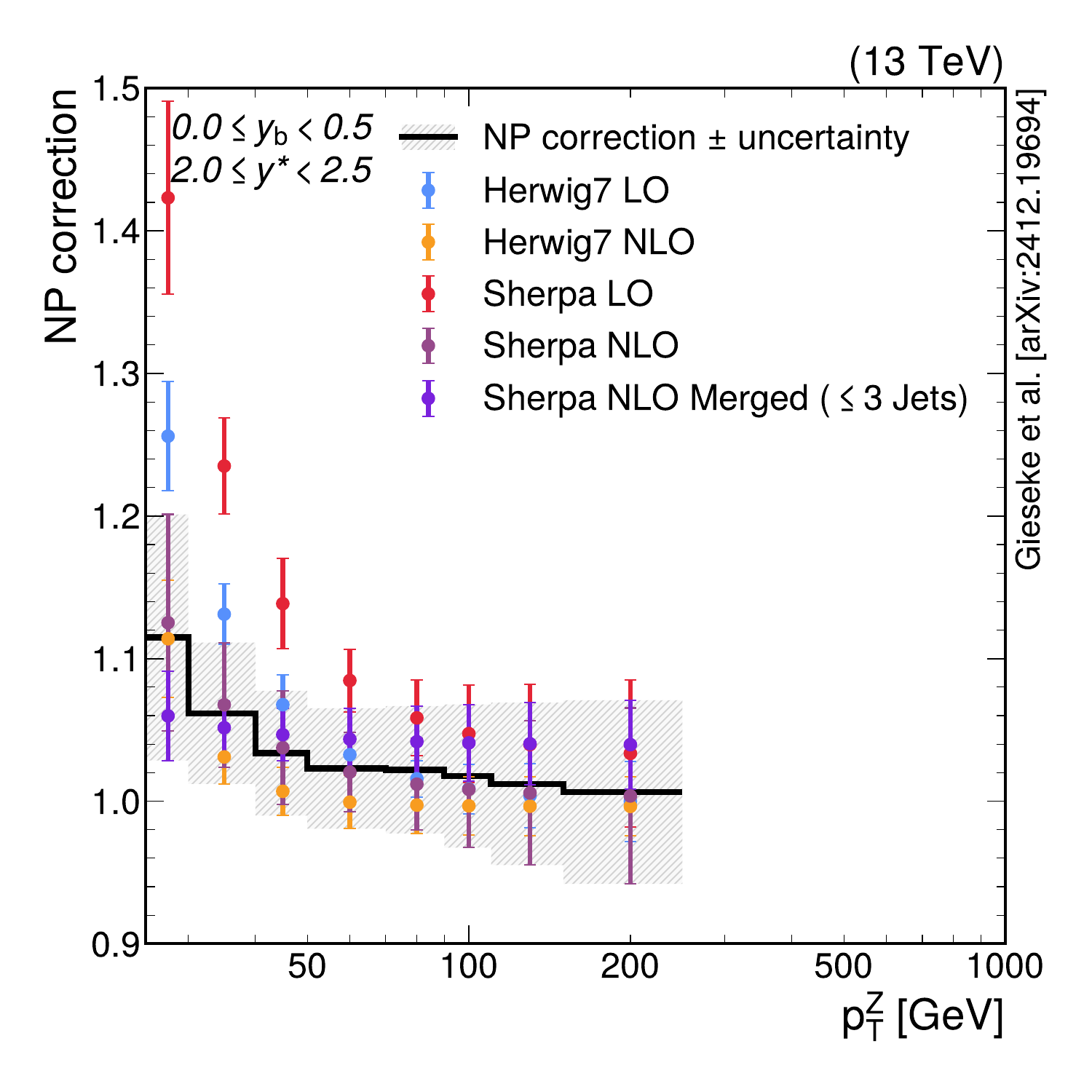}%
  \includegraphics[width=0.5\textwidth]{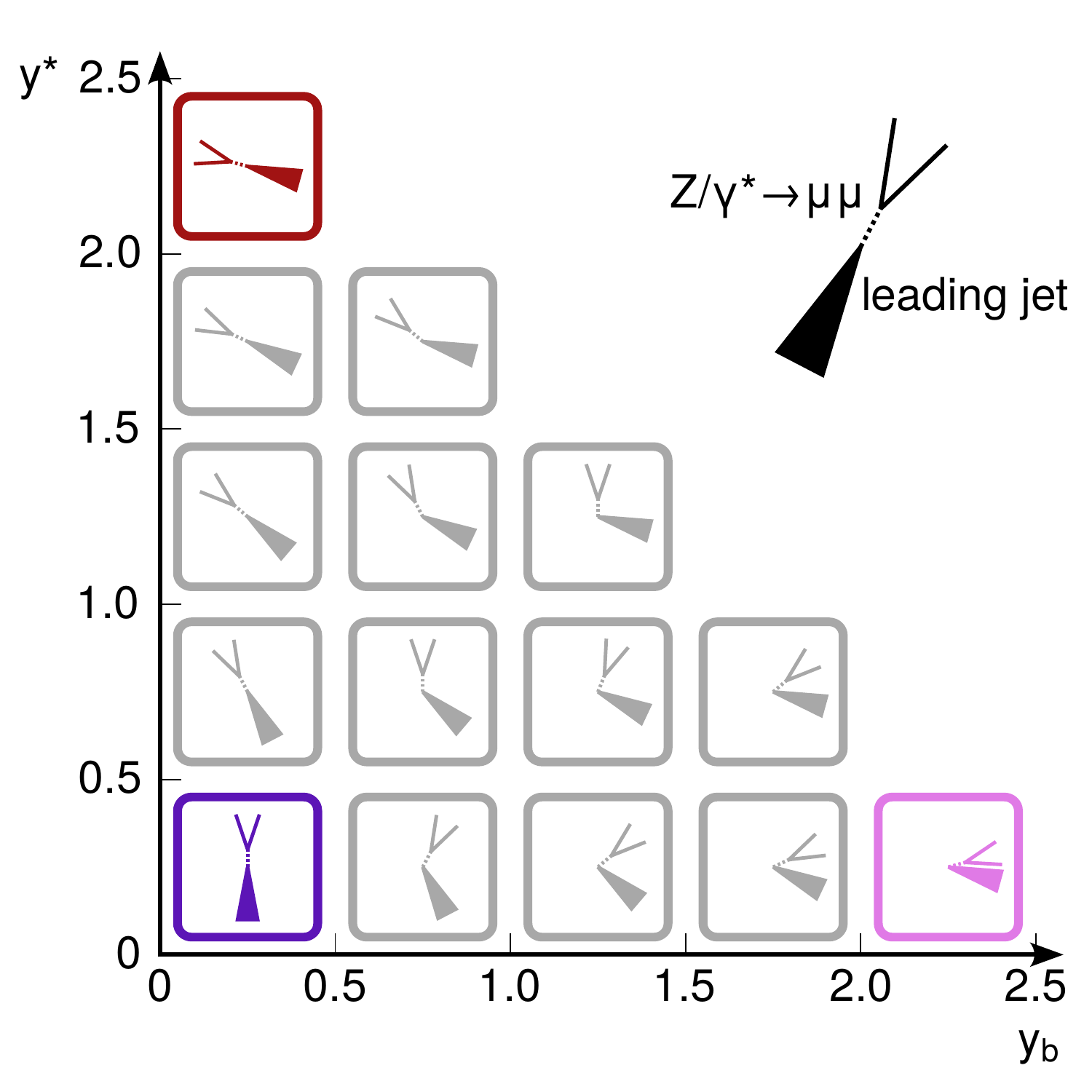}\\
  \includegraphics[width=0.5\textwidth]{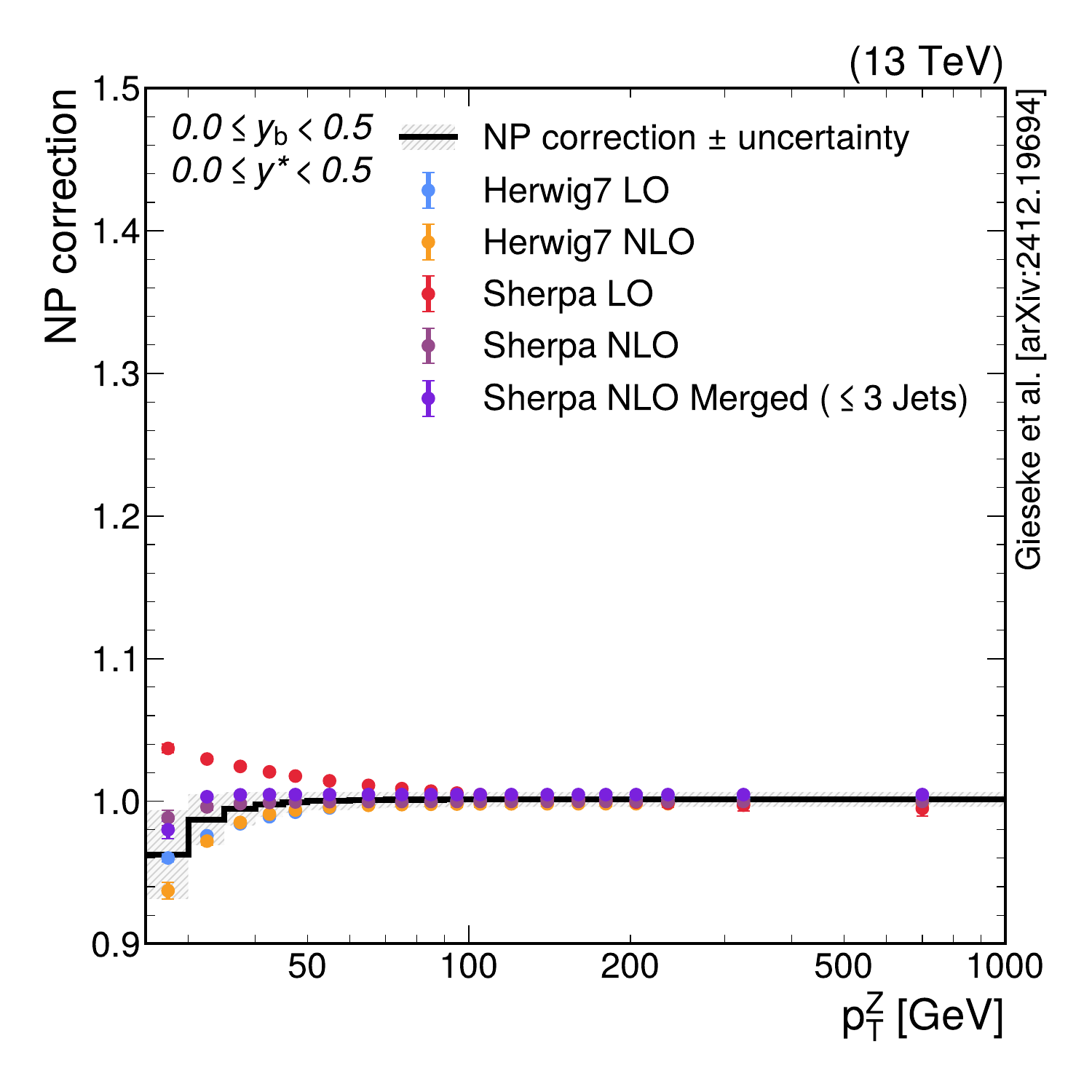}%
  \includegraphics[width=0.5\textwidth]{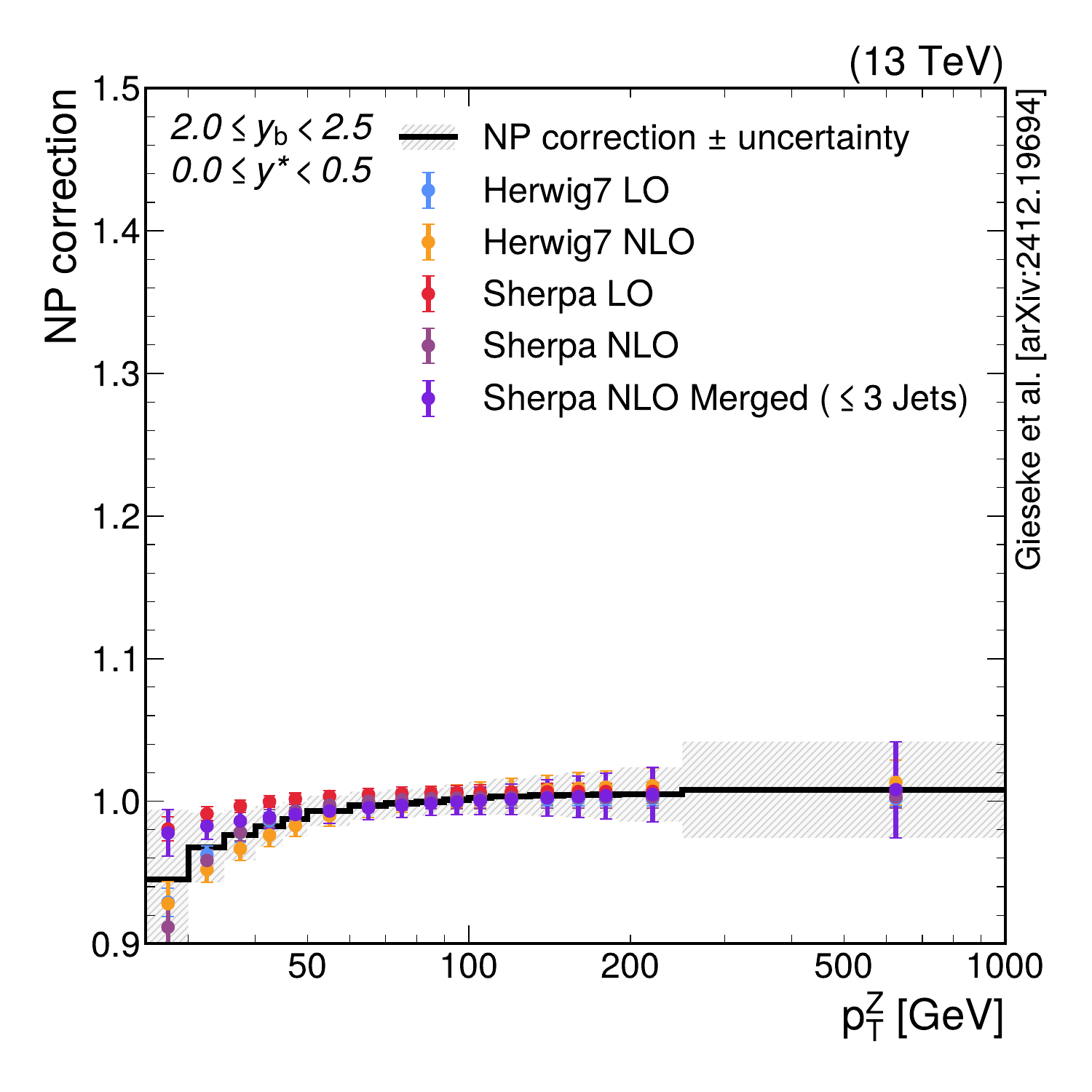}
  \caption{Nonperturbative correction factors obtained from MC simulations using \HERWIGvii or \SHERPA on top of LO or NLO pQCD predictions as
    functions of \ptz for the same three phase space intervals in the \ybys kinematic plane as depicted in Fig.~\ref{fig:uncertainties}.
    An illustration of the presented \ybys bins is shown in the top right.
    The NP correction factors with their respective uncertainties shown as vertical bars are obtained as described in the text from NLO MC simulations.
    For a better visual comparison, the range in \ptz is kept the same for all plots.
  }
  \label{fig:np_corr}
\end{figure}

In order to obtain physical cross sections at the stable-particle level across the whole analyzed phase space, the fixed-order predictions must also be
corrected for NP effects.  These effects become sizable for low \ptz, where \as becomes large and the perturbative
expansion in \as alone is not sufficient anymore.  The NP corrections, using the same triple-differential binning for \zmmj as this analysis, are derived
in Ref.~\cite{Gieseke:2024pqf} using \HERWIG 7.2.3 and \SHERPA 2.2.15.
The NP correction factors are derived by comparing a simulation at the stable-particle level, which includes hadronization and multiparton interactions, to predictions at the parton level based on (N)LO pQCD calculations matched to parton showers.
In Fig.~\ref{fig:np_corr}, the derived correction factors using \HERWIG and \SHERPA with underlying (N)LO pQCD predictions are shown.  Additionally, a
multileg-merged NLO pQCD prediction is used with \SHERPA.

In Ref.~\cite{Gieseke:2024pqf}, it is shown that the NP corrections do not perfectly factorize with the higher-order pQCD
predictions. They are dependent on the underlying pQCD prediction, particularly with respect to changes in \ystar.  Therefore, the final correction is
defined as the envelope comprising the three NP factors, including their statistical uncertainty, as derived from \HERWIGvii and \SHERPA with at least NLO precision in the
baseline pQCD prediction. The center of the envelope is taken as the nominal correction factor with the upwards and downwards spread as uncertainty.
The final correction is shown in Fig.~\ref{fig:np_corr} together with the separate predictions by \HERWIGvii and \SHERPA, and also, for illustration
only, the LO-based NP corrections of the two MC generators.

\section{Comparison of theory predictions to unfolded data}
\label{sec:results}

\begin{figure}[!tp]
  \centering%
  \includegraphics[width=0.5\textwidth]{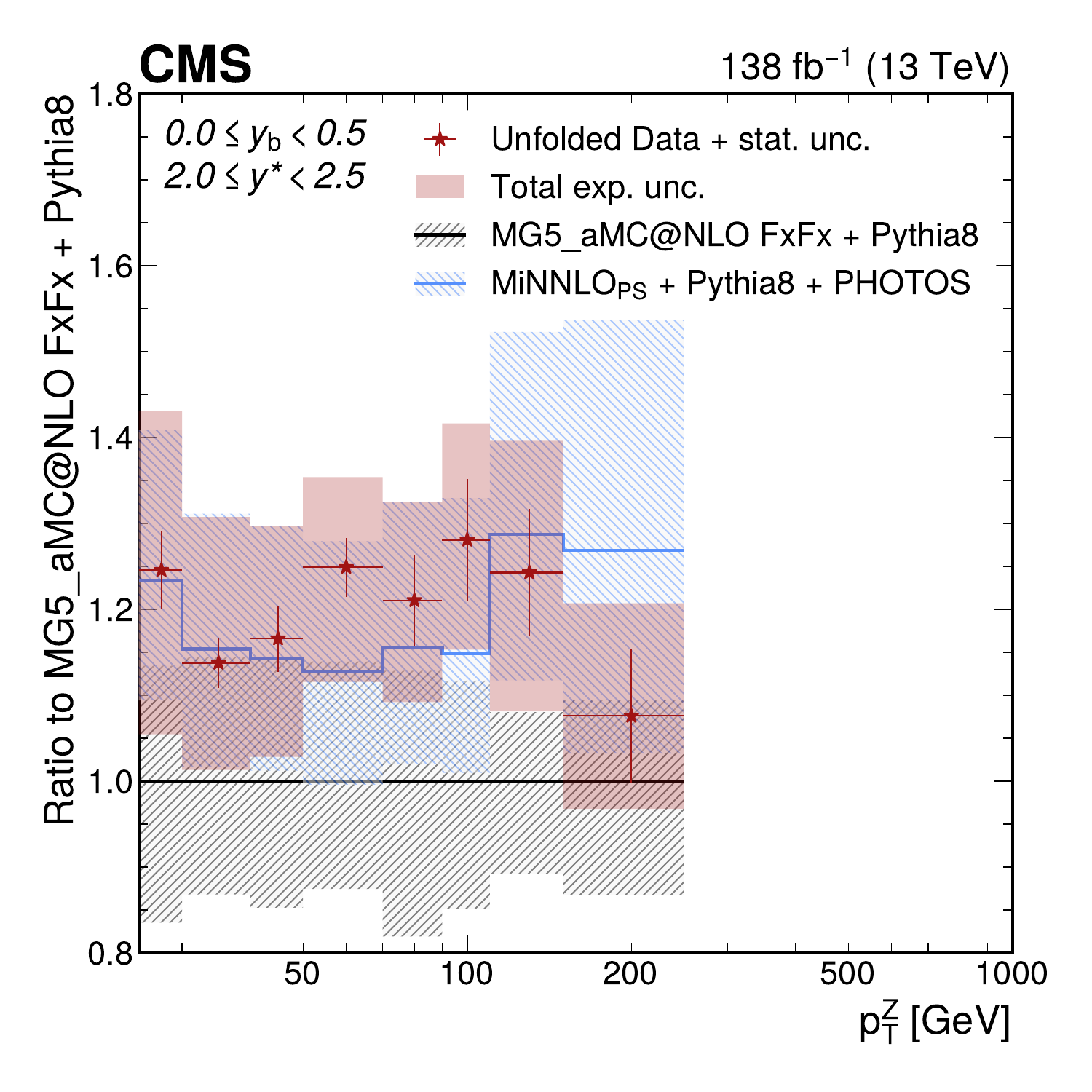}%
  \includegraphics[width=0.5\textwidth]{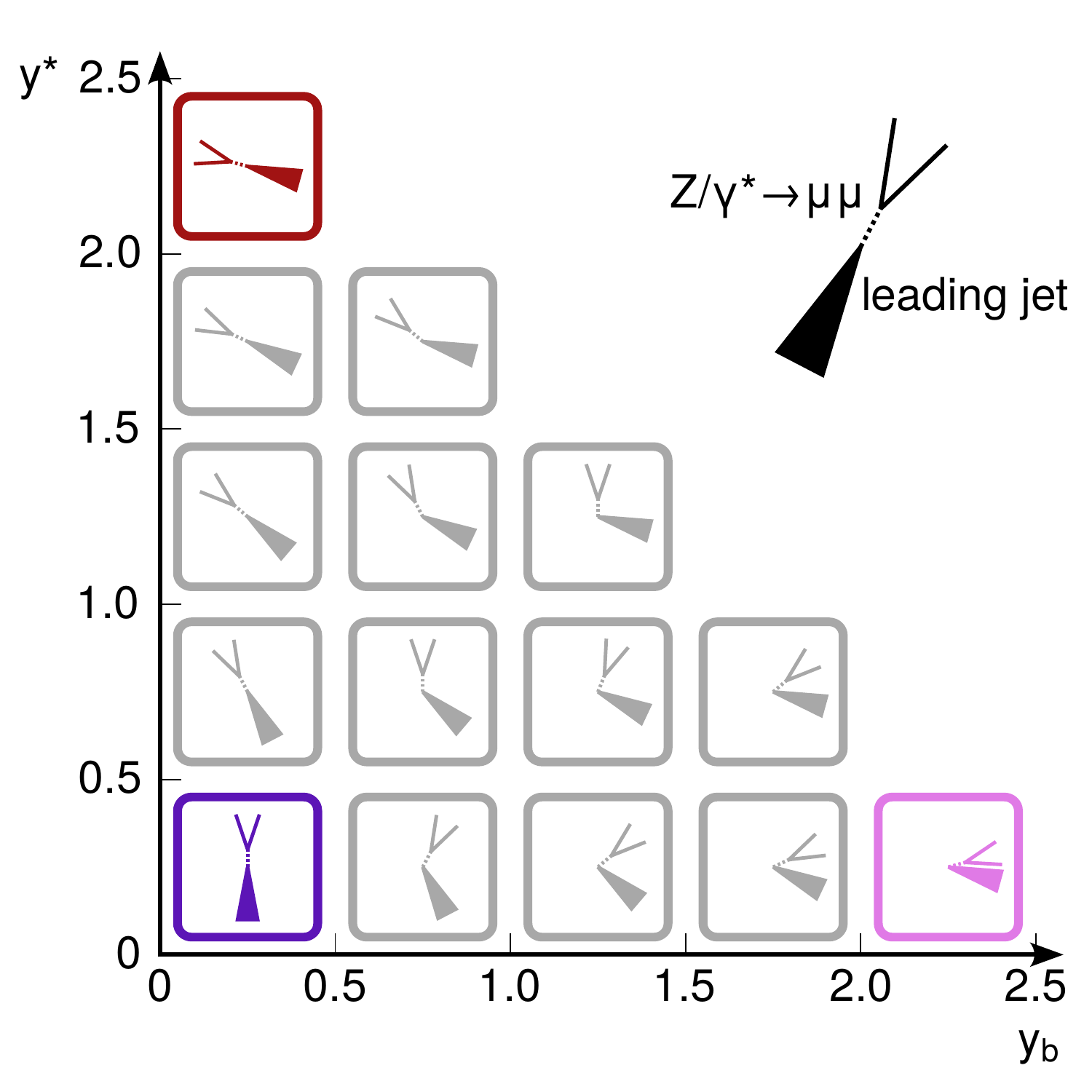}\\
  \includegraphics[width=0.5\textwidth]{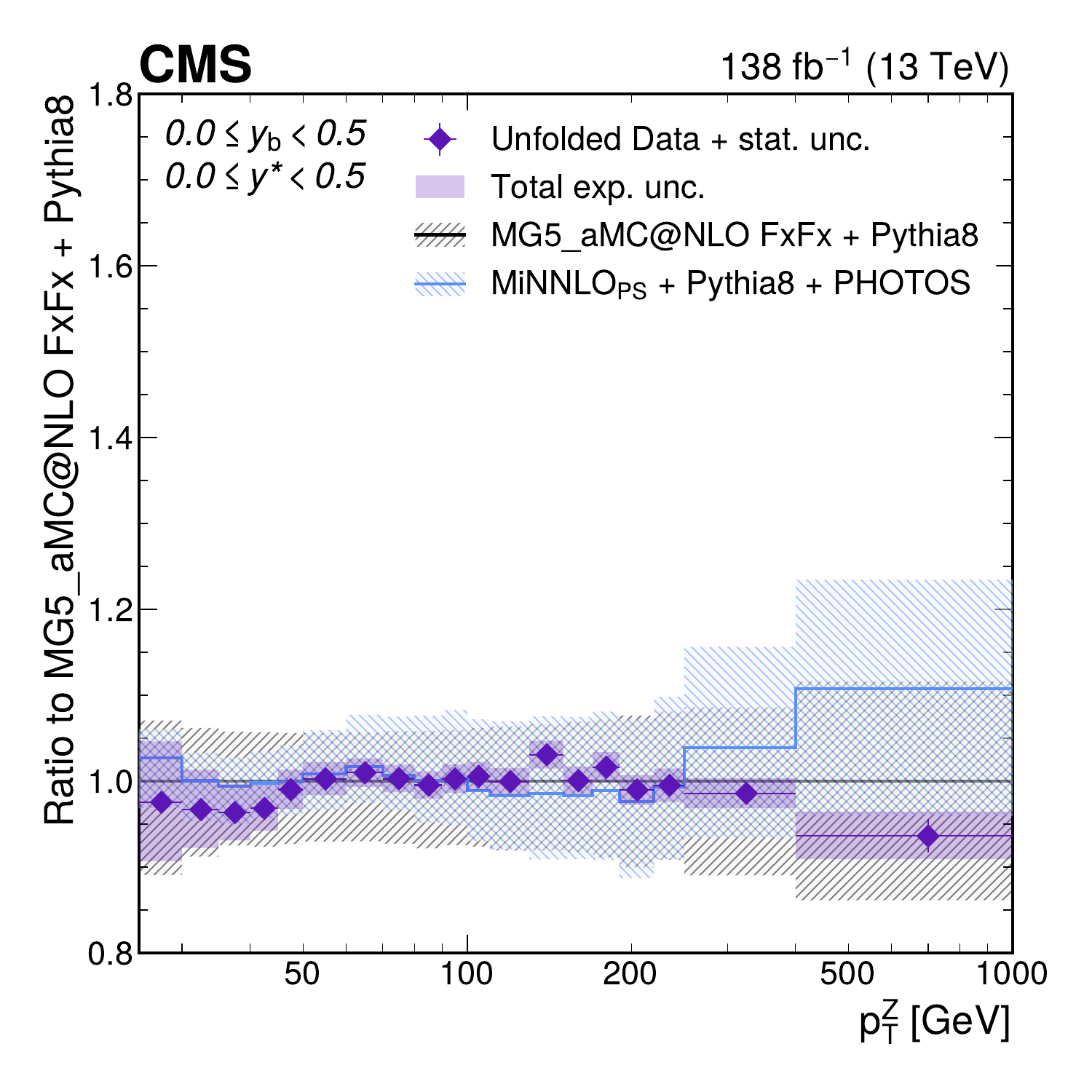}%
  \includegraphics[width=0.5\textwidth]{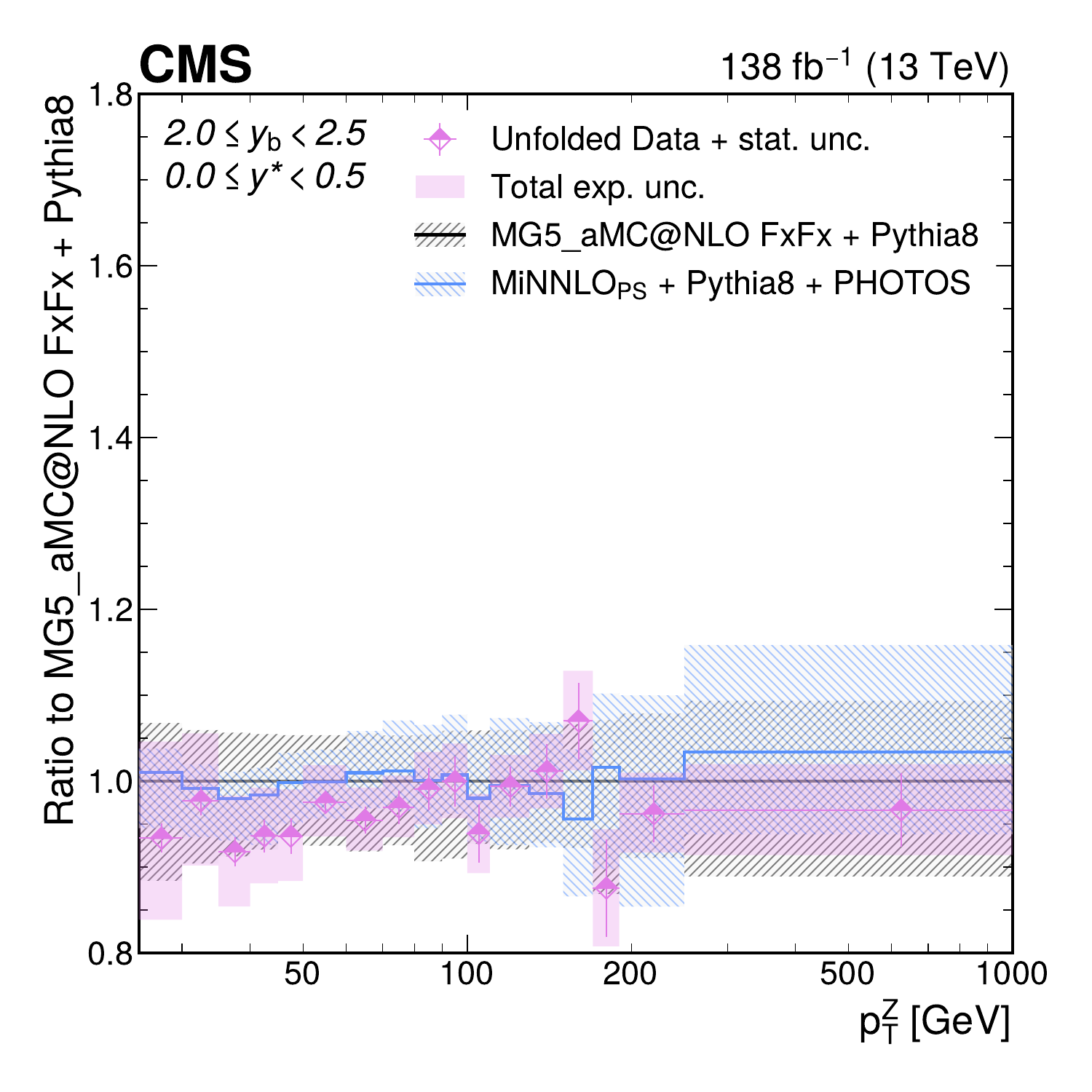}
  \caption{Ratio of the unfolded cross section of \zmmjx production to the MC prediction of \MGvATNLO FxFx + \PYTHIAviii, as a function of \ptz for the same three
    phase space intervals in the \ybys kinematic plane as depicted in Fig.~\ref{fig:uncertainties}.
    An illustration of the presented \ybys bins is shown in the top right.
    The uncertainty band on the MC prediction includes statistical, renormalization and factorization scale, PDF, and initial- and final-state radiation uncertainties. The alternative signal MC sample based on
    \POWHEG{}\,\minnlops + \PYTHIAviii is shown as well.
    The vertical bars on the data points represent the statistical uncertainties, the horizontal bars the bin widths, and the shaded band around the data points the total experimental uncertainty.
    The color and markers of the data points match the representation of the respective \ybys bin as in Figs.~\ref{fig:phase_space} and~\ref{fig:xs_overview}.
    For a better visual comparison, the range in \ptz is kept the same for all plots.
    }\label{fig:xs_mc_comparison}
\end{figure}

The unfolded \zmmjx production cross sections are compared with the predictions at the stable-particle level of \MGvATNLO and \POWHEG{}\,\minnlops corrected for final-state photon radiation with \PHOTOS, both interfaced with \PYTHIAviii.  The ratio to the \MGvATNLO
prediction is shown in Fig.~\ref{fig:xs_mc_comparison} as a function of \ptz for the central bin in \ybys, the extreme \yboost, and extreme \ystar
bins.

The predictions by \MGvATNLO and \POWHEG{}\,\minnlops are both of NLO accuracy in pQCD in the one-jet phase space.
Both use PDF4LHC PDF sets at NLO, but instead of the version PDF4LHC15 for the former, the latter
is based on the newer PDF4LHC21 PDF set.  The statistical and systematic uncertainty in the predictions is shown as
a band that includes scale variations, PDF uncertainties, and uncertainties in the modeling of initial- and final-state radiation, as described in
Section~\ref{sec:uncertainties} for the background simulation.  The error bars on the data points represent the statistical uncertainty, while the
band around the data points indicates the total uncertainty in the measurement.  Overall, a good agreement between data and simulation is observed with
a somewhat better description of the high-\ystar region by \POWHEG{}\,\minnlops + \PYTHIAviii.

\begin{figure}[!tp]
  \centering%
  \includegraphics[width=0.5\textwidth]{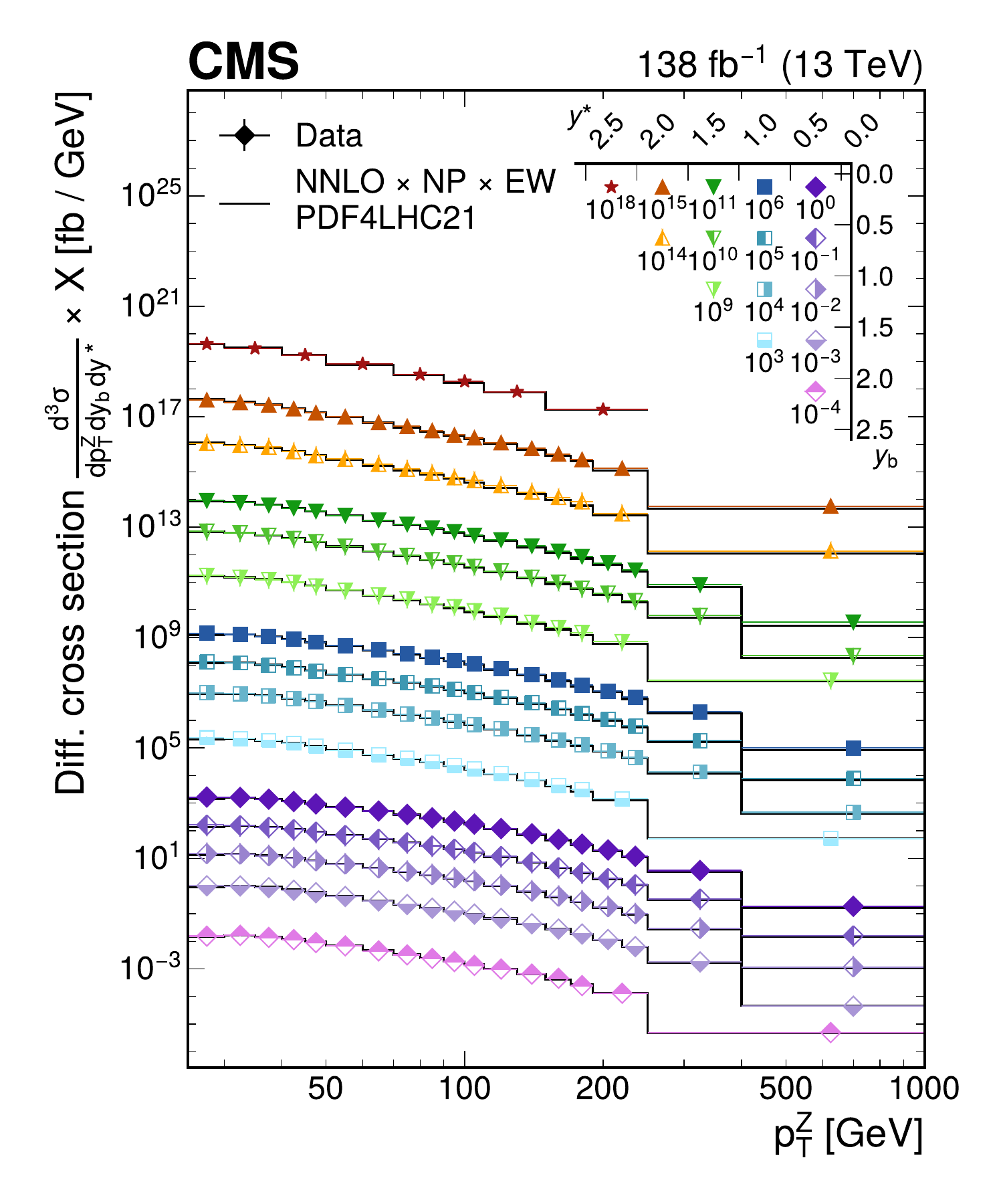}%
  \includegraphics[width=0.5\textwidth]{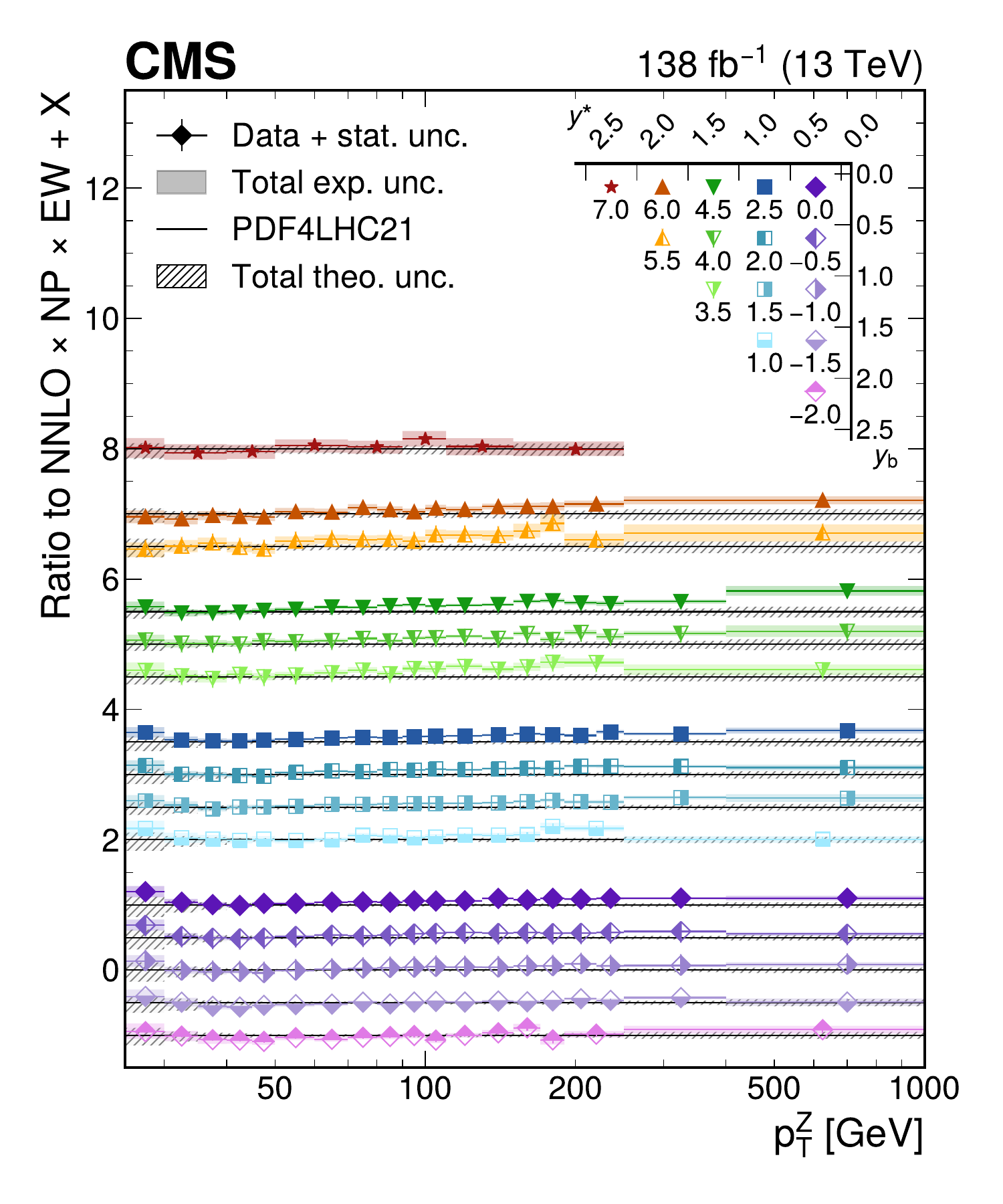}
  \caption{For each \ybys bin, the unfolded cross section of \zmmjx production as a function of \ptz is compared with predictions at NNLO accuracy corrected for EW and NP effects (left). For a better comparison of data versus theory, the right plot shows the ratio of the unfolded cross section to the prediction. For improved visibility, the points and curves are shifted by constant factors (left) or offsets (right) as indicated in the legend.
    The vertical bars on the data points represent the total experimental uncertainty (left) or the statistical uncertainties (right), and the horizontal bars the bin widths. In the left plot, the vertical bars are smaller than the marker size. The shaded bands around the data points (right) indicate the total experimental uncertainty, and the hatched bands the total theoretical uncertainty in the prediction.
    }\label{fig:xs_overview}
\end{figure}

Furthermore, the theory predictions at NNLO accuracy in pQCD provided by \NNLOJet, corrected for NLO EW and NP effects, are
compared in Fig.~\ref{fig:xs_overview} (left) with the unfolded data.
The corresponding ratios of the unfolded data to the NNLO predictions are shown on the right.

\begin{figure}[!tp]
  \centering%
  \includegraphics[width=0.5\textwidth]{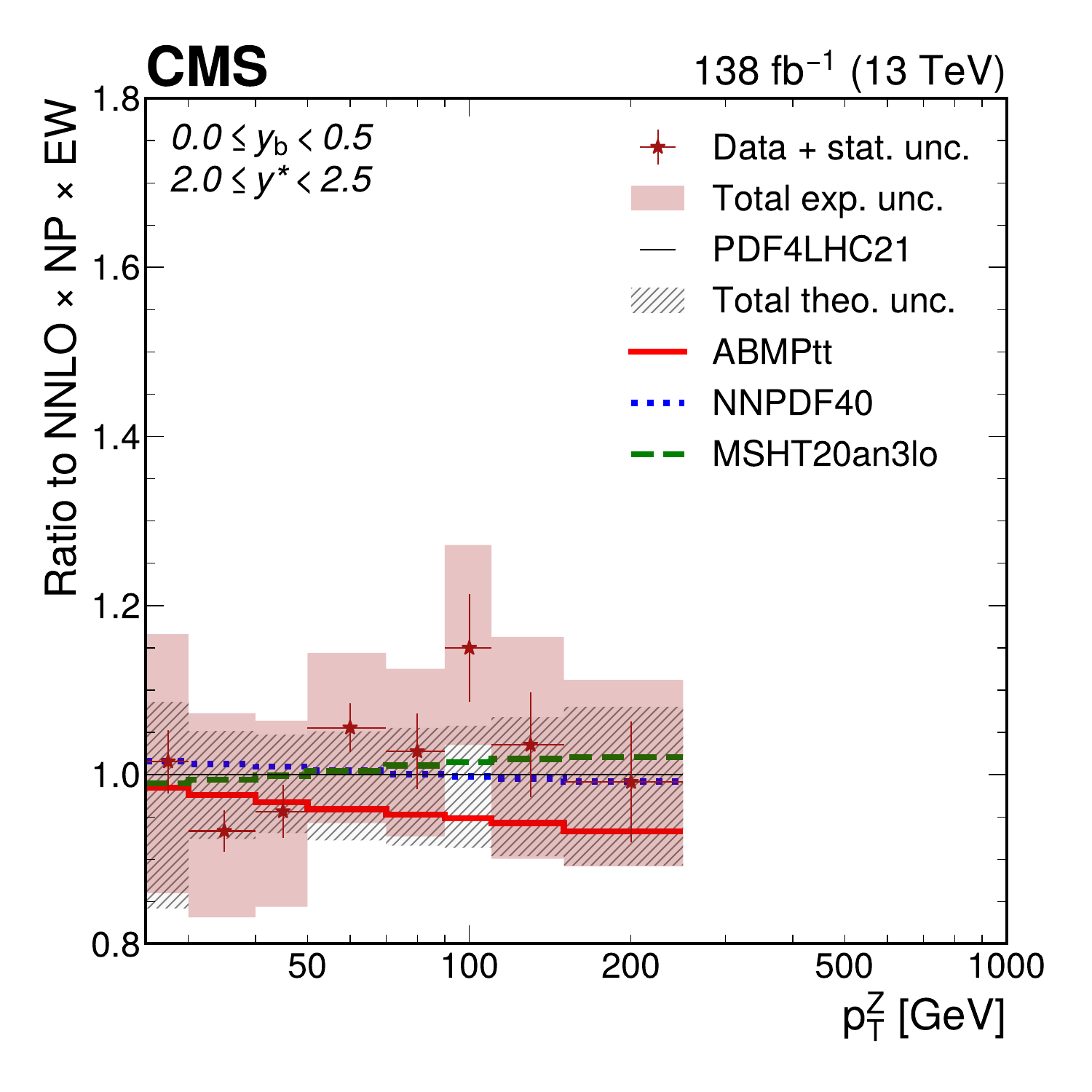}%
  \includegraphics[width=0.5\textwidth]{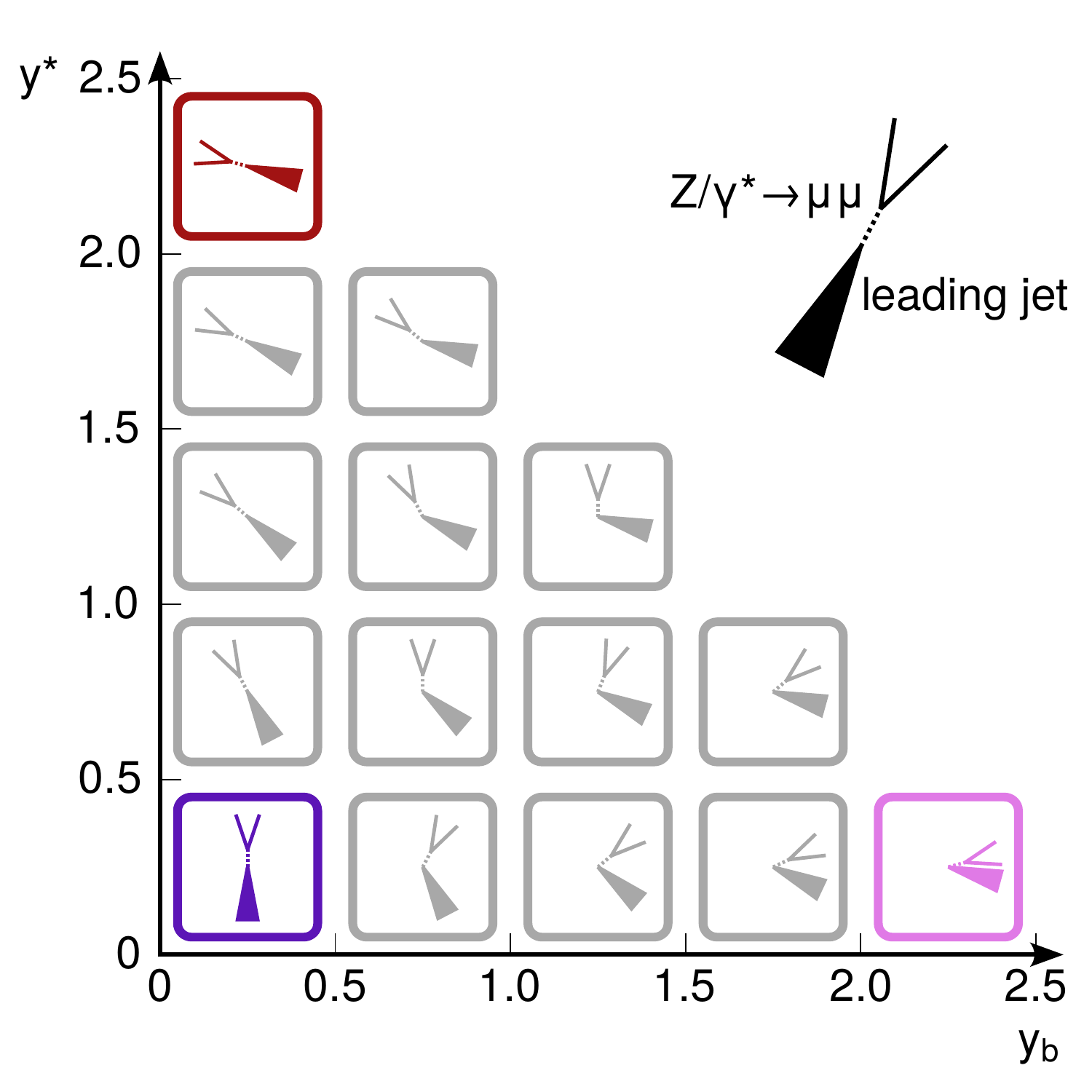}\\
  \includegraphics[width=0.5\textwidth]{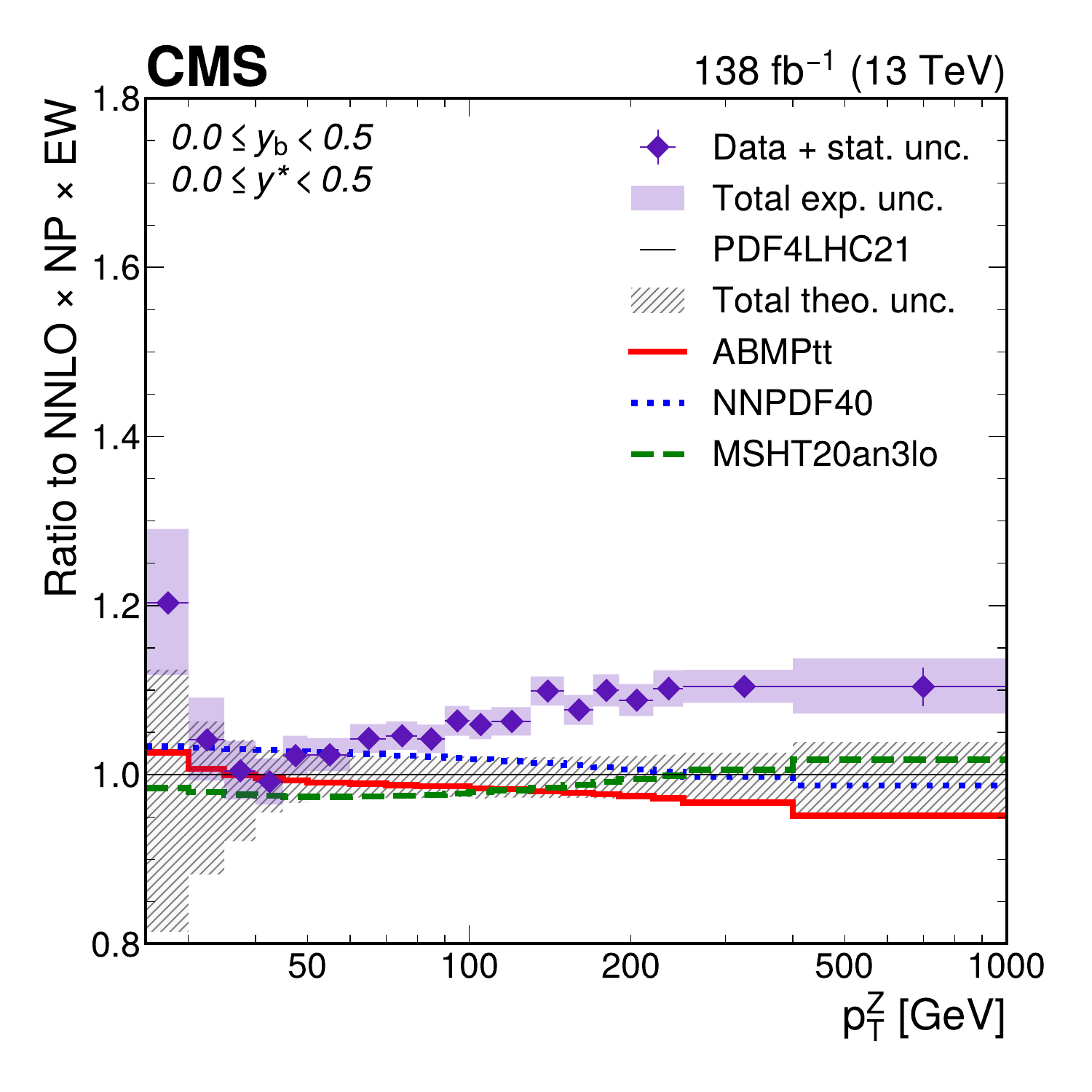}%
  \includegraphics[width=0.5\textwidth]{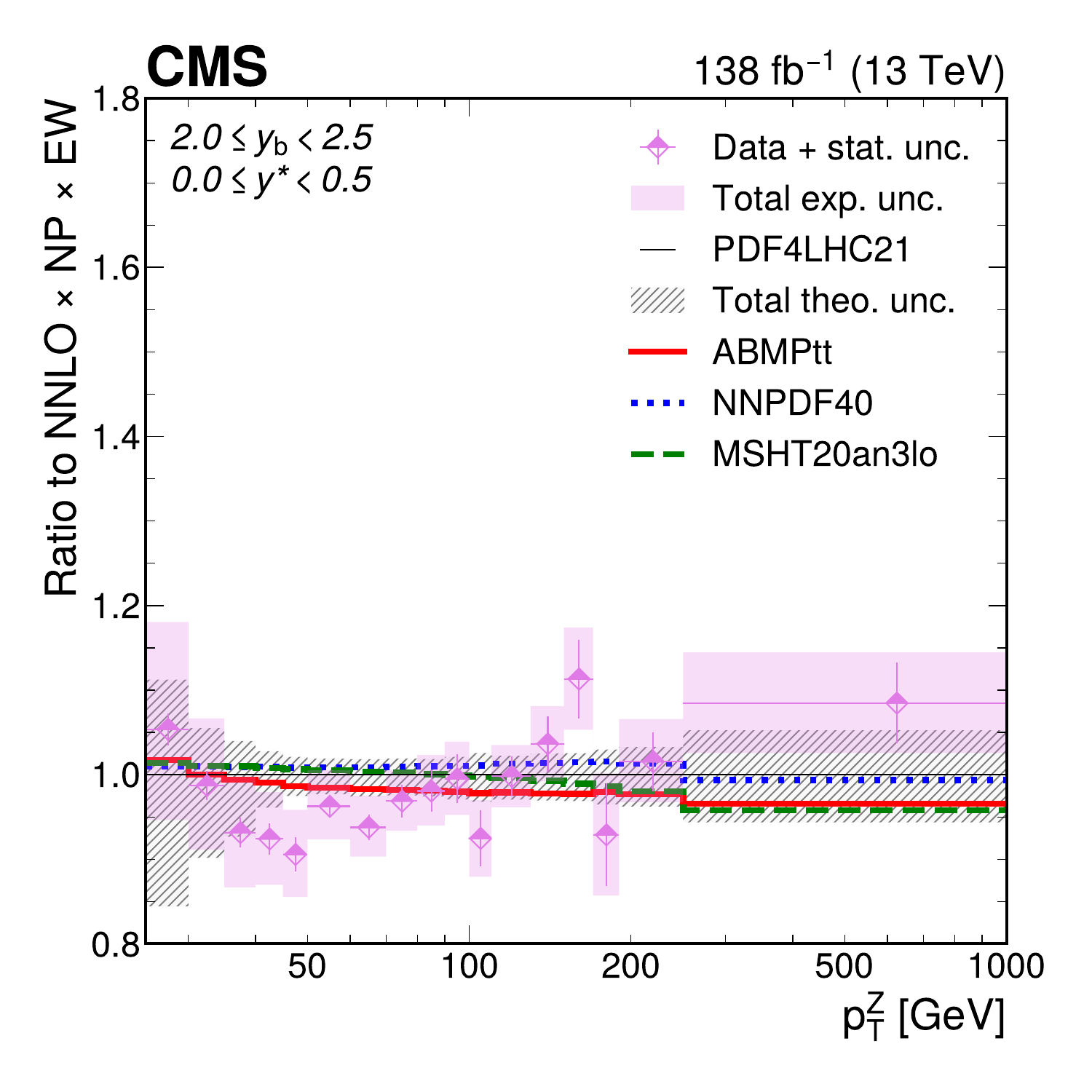}
  \caption{Ratio of the unfolded cross section of \zmmjx production to the NNLO prediction corrected for EW and NP effects, as a function of
    \ptz for the same three phase space intervals in the \ybys kinematic plane as depicted in Fig.~\ref{fig:uncertainties}.
    An illustration of the presented \ybys bins is shown in the top right.
    Predictions using alternative PDF sets are shown as well.
    The vertical bars on the data points represent the statistical uncertainties, the horizontal bars the bin widths, and the shaded band around the data points the total experimental uncertainty.
    The color and markers of the data points match the representation of the respective \ybys bin as in Figs.~\ref{fig:phase_space} and~\ref{fig:xs_overview}.
    For a better visual comparison, the range in \ptz is kept the same for all plots.
  }
  \label{fig:xs_ratios_single}
\end{figure}

Using the central, high-\yboost, and high-\ystar bins as representative examples, Fig.~\ref{fig:xs_ratios_single} shows these ratios also for the alternative PDF sets of:
five-flavor ABMPtt NNLO~\cite{Alekhin:2024bhs}, NNPDF4.0 NNLO~\cite{NNPDF:2021njg}, and MSHT20 approximate next-to-NNLO (aN$^3$LO)~\cite{McGowan:2022nag}.
The predicted cross sections differ among themselves by up to 7\% in the central rapidity region.
The uncertainty bands on the theory predictions include: the renormalization and factorization scale variations, the PDF4LHC21 uncertainties, the uncertainty in
the EW and NP corrections, and the statistical uncertainty in the numerical integration from \NNLOJet.

None of the predictions describes the data within uncertainties across the whole phase space.
Notably, in the central region, where a high experimental precision is achieved, the theory underestimates the measured data by up to 12\%, which is not fully covered by the uncertainties.
In the highly boosted and the forward-backward regions, where the experimental uncertainties are largest, the observed differences become less significant.

\section{Summary}
\label{sec:summary}

A measurement of the triple-differential \zmmjx production cross section is presented, where the data, recorded in the years 2016--2018 and corresponding to an
integrated luminosity of 138\fbinv, have been unfolded for detector effects for the first time simultaneously in three dimensions.  The three observables were
chosen to be the transverse momentum of the \PZ boson, \ptz, as reconstructed from the decay muons, the boost of the center-of-mass system, \yboost,
of the $2\to 2$ scattering process with the \PZ boson and a balancing jet in the final state, and half of their absolute separation in
rapidity, \ystar. These quantities correlate with the energy scale of the collision process, the initial-state parton momenta, and the
scattering angle in the partonic center-of-mass frame, and hence are very advantageous for comparisons to theory.

For $50 < \ptz < 1000\GeV$, the \zmmjx cross section has been measured with a total experimental uncertainty of 1.5\% in the central
detector region, increasing to 3\% up to at most 8\% in the boosted regime of $\yboost > 1$.
In the central region, the uncertainty receives comparable contributions from several systematic effects, whereas at larger \ptz or \yboost it is dominated by limited event counts.
For $\ptz < 50\GeV$, the dominant uncertainty is always given by the jet energy scale that
comes into play through the minimum jet-\pt requirement when defining the leading-\pt jet in the event. Going to larger rapidity separations \ystar,
event numbers drop dramatically and the reach in \ptz is much more limited. The experimental uncertainty, combined from the jet energy scale, the
statistical uncertainties in data and simulation, and the choice of the unfolding model, increases to 10--13\%.

Within the uncertainties, the \zmmjx triple-differential production cross section is well described, in particular by the Monte Carlo event generator
\POWHEG{}\,\minnlops + \PYTHIAviii, which is of next-to-leading order accuracy only in perturbative quantum chromodynamics when requiring at least one jet.
In comparison with accurate predictions at next-to-next-to-leading order, complemented with electroweak and
nonperturbative corrections, discrepancies of up to 12\% are found notably in the central rapidity region.
While alternative parton distribution function sets vary the cross section by up to 7\% there, none of the predictions fully describes the data.
This demonstrates the potential of these data to further constrain the partonic structure of the proton.
At larger \yboost and \ystar, where the experimental uncertainties are largest, the observed differences become less significant.
The predicted decrease in cross section for larger \ystar is confirmed by data and can be understood in terms of the parton distributions in the proton that disfavor this kinematic regime for parton-parton processes available at leading order. Only at next-to-leading order, quark-quark scattering can contribute to this cross section.
This parton-parton subprocess decomposition in the triple-differential \zj production, which differs significantly from the one in dijet production, is very promising with regard to future determinations of
parton distributions and the strong coupling in combined fits together with LHC dijet data.

\begin{acknowledgments}

We congratulate our colleagues in the CERN accelerator departments for the excellent performance of the LHC and thank the technical and administrative staffs at CERN and at other CMS institutes for their contributions to the success of the CMS effort. In addition, we gratefully acknowledge the computing centers and personnel of the Worldwide LHC Computing Grid and other centers for delivering so effectively the computing infrastructure essential to our analyses. Finally, we acknowledge the enduring support for the construction and operation of the LHC, the CMS detector, and the supporting computing infrastructure provided by the following funding agencies: SC (Armenia), BMFWF and FWF (Austria); FNRS and FWO (Belgium); CNPq, CAPES, FAPERJ, FAPERGS, and FAPESP (Brazil); MES and BNSF (Bulgaria); CERN; CAS, MoST, and NSFC (China); MINCIENCIAS (Colombia); MSES and CSF (Croatia); RIF (Cyprus); SENESCYT (Ecuador); ERC PRG and PSG, TARISTU24-TK10 and MoER TK202 (Estonia); Academy of Finland, MEC, and HIP (Finland); CEA and CNRS/IN2P3 (France); SRNSF (Georgia); BMFTR, DFG, and HGF (Germany); GSRI (Greece); MATE and NKFIH (Hungary); DAE and DST (India); IPM (Iran); SFI (Ireland); INFN (Italy); MSIT and NRF (Republic of Korea); MES (Latvia); LMTLT (Lithuania); MOE and UM (Malaysia); BUAP, CINVESTAV, CONACYT, LNS, SEP, and UASLP-FAI (Mexico); MOS (Montenegro); MBIE (New Zealand); PAEC (Pakistan); MSHE, NSC, and NAWA (Poland); FCT (Portugal);  MESTD (Serbia); MICIU/AEI and PCTI (Spain); MOSTR (Sri Lanka); Swiss Funding Agencies (Switzerland); NSTC (Taiwan); MHESI (Thailand); TUBITAK and TENMAK (T\"{u}rkiye); NASU (Ukraine); STFC (United Kingdom); DOE and NSF (USA).

\begin{sloppypar}
\setlength\emergencystretch{\hsize}
\hyphenation{Rachada-pisek} Individuals have received support from the Marie-Curie program and the European Research Council and Horizon 2020 Grant, contract Nos.\ 675440, 724704, 752730, 758316, 765710, 824093, 101115353, 101002207, 101001205, and COST Action CA16108 (European Union); the Leventis Foundation; the Alfred P.\ Sloan Foundation; the Alexander von Humboldt Foundation; the Science Committee, project no. 22rl-037 (Armenia); the Fonds pour la Formation \`a la Recherche dans l'Industrie et dans l'Agriculture (FRIA) and Fonds voor Wetenschappelijk Onderzoek contract No. 1228724N (Belgium); the Beijing Municipal Science \& Technology Commission, No. Z191100007219010, the Fundamental Research Funds for the Central Universities, the Ministry of Science and Technology of China under Grant No. 2023YFA1605804, the Natural Science Foundation of China under Grant No. 12535004, and USTC Research Funds of the Double First-Class Initiative No.\ YD2030002017 (China); the Ministry of Education, Youth and Sports (MEYS) of the Czech Republic; the Shota Rustaveli National Science Foundation (Georgia); the Deutsche Forschungsgemeinschaft (DFG), among others, under Germany's Excellence Strategy -- EXC 2121 ``Quantum Universe" -- 390833306, and under project numbers 400140256 - GRK2497 and INST 39/963-1 FUGG,  the state of Baden-W\"urttemberg through bwHPC; the Hellenic Foundation for Research and Innovation (HFRI), Project Number 2288 (Greece); the Hungarian Academy of Sciences, the New National Excellence Program - \'UNKP, the NKFIH research grants K 131991, K 138136, K 143460, K 143477, K 147557, K 146913, K 146914, K 147048, TKP2021-NKTA-64, 2026-4.1.1-MISSZI\'{O}K-2026-00005 and 2025-1.1.5-NEMZ\_KI-2025-00004, and MATE KKP and KKPCs Research Excellence and Flagship Research Groups grants (Hungary); the Council of Science and Industrial Research, India; ICSC -- National Research Center for High Performance Computing, Big Data and Quantum Computing, FAIR -- Future Artificial Intelligence Research, and CUP I53D23001070006 (Mission 4 Component 1), funded by the NextGenerationEU program, the Italian Ministry of University and Research (MUR) under Bando PRIN 2022 -- CUP I53C24002390006, PRIN PRIMULA 2022RBYK7T (Italy); the Latvian Council of Science; the Ministry of Science and Higher Education, project no. 2022/WK/14, and the National Science Center, contracts Opus 2021/41/B/ST2/01369, 2021/43/B/ST2/01552, 2023/49/B/ST2/03273, and the NAWA contract BPN/PPO/2021/1/00011 (Poland); the Funda\c{c}\~ao para a Ci\^encia e a Tecnologia (Portugal); the National Priorities Research Program by Qatar National Research Fund; MICIU/AEI/10.13039/501100011033, ERDF/EU, ``European Union NextGenerationEU/PRTR", projects PID2022-142604OB-C21, PID2022-139519OB-C21, PID2023-147706NB-I00, PID2023-148896NB-I00, PID2023-146983NB-I00, PID2023-147115NB-I00, PID2023-148418NB-C41, PID2023-148418NB-C42, PID2023-148418NB-C43, PID2023-148418NB-C44, PID2024-158190NB-C22, RYC2021-033305-I, RYC2024-048719-I, CNS2023-144781, CNS2024-154769 and Plan de Ciencia, Tecnolog{\'i}a e Innovaci{\'o}n de Asturias, Spain; the Chulalongkorn Academic into Its 2nd Century Project Advancement Project, the National Science, Research and Innovation Fund program IND\_FF\_68\_369\_2300\_097, and the Program Management Unit for Human Resources \& Institutional Development, Research and Innovation, grant B39G680009 (Thailand); the Eric \& Wendy Schmidt Fund for Strategic Innovation through the CERN Next Generation Triggers project under grant agreement number SIF-2023-004; the Kavli Foundation; the Nvidia Corporation; the SuperMicro Corporation; the Welch Foundation, contract C-1845; and the Weston Havens Foundation (USA).
\end{sloppypar}
\end{acknowledgments}\section*{Data availability} Release and preservation of data used by the CMS Collaboration as the basis for publications is guided by the  \href{https://doi.org/10.7483/OPENDATA.CMS.1BNU.8V1W}{CMS data preservation, re-use and open access policy}.

\bibliography{auto_generated}
\cleardoublepage \appendix\section{The CMS Collaboration \label{app:collab}}\begin{sloppypar}\hyphenpenalty=5000\widowpenalty=500\clubpenalty=5000\cmsinstitute{Yerevan Physics Institute, Yerevan, Armenia}
{\tolerance=6000
A.~Gevorgyan\cmsorcid{0000-0003-2751-9489}, A.~Hayrapetyan, V.~Makarenko\cmsorcid{0000-0002-8406-8605}, A.~Tumasyan\cmsAuthorMark{1}\cmsorcid{0009-0000-0684-6742}
\par}
\cmsinstitute{Marietta Blau Institute for Particle Physics, Vienna, Austria}
{\tolerance=6000
W.~Adam\cmsorcid{0000-0001-9099-4341}, L.~Benato\cmsorcid{0000-0001-5135-7489}, T.~Bergauer\cmsorcid{0000-0002-5786-0293}, M.~Dragicevic\cmsorcid{0000-0003-1967-6783}, S.~Gundacker\cmsorcid{0000-0003-2087-3266}, A.K.~Guven\cmsorcid{0009-0004-5670-5138}, P.S.~Hussain\cmsorcid{0000-0002-4825-5278}, M.~Jeitler\cmsAuthorMark{2}\cmsorcid{0000-0002-5141-9560}, N.~Krammer\cmsorcid{0000-0002-0548-0985}, A.~Li\cmsorcid{0000-0002-4547-116X}, D.~Liko\cmsorcid{0000-0002-3380-473X}, M.~Matthewman, J.~Schieck\cmsAuthorMark{2}\cmsorcid{0000-0002-1058-8093}, R.~Sch\"{o}fbeck\cmsAuthorMark{2}\cmsorcid{0000-0002-2332-8784}, M.~Shooshtari\cmsorcid{0009-0004-8882-4887}, M.~Sonawane\cmsorcid{0000-0003-0510-7010}, N.~Van~Den~Bossche\cmsorcid{0000-0003-2973-4991}, W.~Waltenberger\cmsorcid{0000-0002-6215-7228}, C.E.~Wulz\cmsAuthorMark{2}\cmsorcid{0000-0001-9226-5812}
% ...
\par}
$^{1}$Also at Yerevan State University, Yerevan, Armenia\\
$^{2}$Also at Technische Universit\"{a}t Wien, Vienna, Austria\\
% ...
\end{sloppypar}
\end{document}